\documentclass{svjour2}                    
\smartqed  
\usepackage{graphicx}
\usepackage{color}
\begin{document}

\title{Kenneth Wilson and lattice QCD
}


\author{Akira Ukawa         
}


\institute{RIKEN Advanced Institute for Computational Science \at
              7-1-26 Minami-machi, Minatojima, Chuoku, Kobe 650-0047, Japan \\
              Tel.: +81 78 940 5575\\
              \email{akira.ukawa@riken.jp}           
}

\date{Received: date / Accepted: date}

\maketitle

\begin{abstract}
We discuss the physics and computation of lattice QCD, a space-time lattice formulation of quantum chromodynamics, and Kenneth Wilson's seminal role in its development.  
We start with the fundamental issue of confinement of quarks in the theory of the strong interactions, and discuss how lattice QCD provides a framework for understanding this phenomenon.   A conceptual issue with lattice QCD is  a conflict of space-time lattice with chiral symmetry of quarks.  We discuss how this problem is resolved.  Since lattice QCD is a non-linear quantum dynamical system with infinite degrees of freedom, quantities which are analytically calculable are limited.  On the other hand, it provides an ideal case of massively parallel numerical computations.  We review the long and distinguished history of parallel-architecture supercomputers designed and built for lattice QCD.  We discuss algorithmic developments, in particular the difficulties posed by the fermionic nature of quarks, and their resolution.   The triad of efforts toward better understanding of physics, better algorithms, and  more powerful supercomputers have produced major breakthroughs in our understanding of the strong interactions.   We review the salient results of this effort in understanding the hadron spectrum, the Cabibbo-Kobayashi-Maskawa matrix elements and CP violation, and quark-gluon plasma at high temperatures.  We conclude with a brief summary and a future perspective. 

\end{abstract}

\section{Introduction}
\label{intro}

In early 1974, Kenneth Wilson circulated a preprint entitled "Confinement of Quarks".  The paper was received by Physical Review D on 12 June 1974, and was published in the 15 October issue of that journal~\cite{Wilson1974}.  In this paper, Wilson formulated gauge theories on a space-time lattice\footnote{Contemporary research toward lattice gauge theory  was described by Wilson in his plenary talk at 1983 Lepton Photon Symposium~\cite{Wilson1983}, and in more detail in his historical talk at 2004 International Symposium on Lattice Field Theory at FNAL~\cite{Wilson2005}.}.  Using an expansion in inverse powers of the bare gauge coupling constant, Wilson demonstrated that lattice gauge theories at strong coupling confine charged states.  He also argued that the absence of Lorentz invariance (or Euclidean invariance for the imaginary time used for the lattice formulation) is not a hindrance if there is a second order phase transition at some value of the gauge coupling constant, for in the vicinity of such a phase transition the lattice spacing can be taken to zero while fixing the physical correlation length at a finite value.  

This paper laid the conceptual foundation for understanding the quark confinement phenomenon.  It showed that large quantum fluctuations of gauge fields at strong coupling can generate a force between charged states which stays constant at arbitrary distances.  This is a novel type of force, essentially different from the Yukawa force due to exchange of particles which tends to zero at large distances.  

Initially, Wilson's idea did not catch on rapidly.  Techniques were hard to come by which allowed calculations of physical quantities such as hadron masses and connect them to physical predictions in the continuum space-time.  The situation changed dramatically around 1979--1980 when Creutz, Jacobs and Rebbi~\cite{CreutzJacobsRebbi1979},  and Wilson himself~\cite{Wilson1979}, showed the possibility of numerically calculating the observables on a computer.  Particularly dramatic was the calculation of the static quark-antiquark potential by Creutz for SU(2) gauge group in 1979~\cite{Creutz1979}, and the calculations in 1981 of hadron masses by Weingarten~\cite{Weingarten1982} and by Hamber and Parisi~\cite{HamberParisi1981}.  

Traditional quantum field theory until that time could only deal with weakly coupled bound states such as hydrogen.  The possibility of a technique which enables calculation of the properties of relativistic and strongly coupled bound states such as pion and proton was entirely new.  

The timing was also perfect from computational point of view.  The CRAY-1 supercomputer which appeared in 1976 had revolutionized scientific computing, and lattice QCD computation could quickly exploit vector supercomputers  in the 80's.  Perhaps more important in retrospect, rapid development of microprocessors in the 70's stimulated more than a few groups of particle theorists  around the world to start developing parallel computers for lattice QCD.  

The development of lattice QCD has been continuous since then.  With large-scale numerical simulations on parallel supercomputers, understanding of physics of lattice QCD progressed, which in turn led to better algorithms for computation.  These algorithms allowed a better exploitation of the next generation of more powerful computers, which brought even more progress of physics.  

In the four decades of progress, lattice QCD has brought a deep understanding on the physics of the strong interactions.  It has matured to the point where one can make calculations with  the physical values of quark masses,  on lattices with sufficiently large sizes,  at  lattice spacings small enough so that a continuum limit can be carried out with confidence.  

In this article we review lattice QCD from four perspectives in the following four chapters.   In Chapter~\ref{sec:1}, we discuss the foundation of lattice QCD, touching upon the connection between quantum field theory in Euclidean space time and statistical mechanics which was consciously exploited by Wilson.  We show how it led to a conceptual breakthrough in the understanding of confinement. We also explain the issues related with chiral symmetry.  In Chapter~\ref{sec:2}, we discuss the computational aspects.  We review how lattice QCD embodies an ideal case of massively parallel computation, and how this led to the development of parallel supercomputers for lattice QCD, which impacted seriously  on the history of supercomputers up to the present time.  Also discussed is a special computational difficulty posed by the fermionic nature of quark fields, and how overcoming that difficulty has led to the algorithm in use today.  In Chapter~\ref{sec:3}  we discuss some major physics results achieved so far in lattice QCD.  The themes include the hadron mass spectrum, the determination of the Cabibbo-Kobayashi-Maskawa matrix and CP violation in the Standard Model, and the properties of quark gluon plasma at high temperatures and densities.  Finally, in Chapter~\ref{sec:4} we collect some thoughts on how Kenneth Wilson's thinking and vision helped develop the subject.   

The tone of this article is partly historical, describing the development of lattice QCD and the role Kenneth Wilson played in it.  It also reviews  the achievements made in the four decades since its inception in 1974.   

\section{Quantum chromodynamics on a space-time lattice}
\label{sec:1}

\subsection{Hadrons, Quarks, Quantum Chromodynamics}
\label{sec:1-1}

If one looks up the Reviews of Particle Physics web page~\cite{PDG}, one finds  there an entire list of particles and their properties experimentally discovered to date.  In addition to ``Gauge and Higgs Bosons", ``Leptons" and ``Quarks", there are two lists named ``Mesons" and ``Baryons", each of which contains hundreds of particles.  Protons and neutrons, which make up atomic nuclei, are two representative particles belonging to the family of baryons.  Pions and K mesons are less familiar, but important particles for binding protons and neutrons into nuclei, and they belong to the family of mesons.  The mesons and baryons are collectively called ``hadrons".  Their chief characteristic is that they participate in the strong interactions in addition to the electromagnetic and weak interactions, while leptons participate only in the latter two interactions.  

Many of hadrons were discovered in the accelerator experiments in the 50's and 60's.   
In 1964 Gell-Mann and Zweig proposed that hadrons are composed of  more fundamental particles, which Gell-Mann named quarks.  Quarks were predicted to have unusual properties such as fractional charge in units of electron charge.  Evidence has gradually built up, however,  that quarks are real entities.  Yet experimental efforts for detecting them in isolation have been unsuccessful.  This situation is often called ``quark confinement".

Quantum chromodynamics (QCD) proposes to explain the constitution of hadrons from quarks and their  interactions. It is a  quantum field theory with local SU(3) gauge invariance in which  the quark field $q(x)$,  with $x$ the space-time coordinate,  transforms under the fundamental ${\bf 3}$ representation of SU(3).  The SU(3) quantum numbers are called color since the basic premise of the theory is that only color neutral states, trivial under SU(3), carry finite energy and hence exist as physical states.  This is a general statement of "quark confinement".  

Local gauge invariance is a requirement that the frame of reference of an internal symmetry may be freely rotated at each space-time point without altering the content of the theory.  
Thus QCD as a gauge theory  is to remain invariant under the transformation $q(x)\to V(x)q(x)$ where $V(x)\in {\rm SU(3)}$ may vary from point to point in space-time.  This invariance requires the existence of a vector gluon field $A_\mu(x)$ with values in the Lie algebra of SU(3) which tells how the local frame at a point $x$ and at a different point $y$ are related.  

QCD as field theory thus contains gluon as well as quark fields.  It is defined by the Lagrangian density given by 
\begin{eqnarray}
\label{eq:QCDLagrangian}
\cal{L}_{\rm QCD}&=&\frac{1}{2g^2}{\rm Tr}\left(F_{\mu\nu}(x)^2\right)+\overline{q}(x)\left(i\gamma_\mu D_\mu+m\right)q(x).
\end{eqnarray}
Here, $q (x)=(q_1(x), \cdots, q_{N_f})$ is a spin 1/2 Dirac spinor field for quarks, $N_f$ denotes the number of quark flavors with $m=(m_1, \cdots, m_{N_f})$ the quark mass matrix,  and $D_\mu=\partial_\mu-iA_\mu (x)$ is the covariant derivative with  $F_{\mu\nu}=\partial_\mu A_\nu-\partial_\nu A_\mu+[A_\mu, A_\nu]$ the gluon field strength, and $g$ is the QCD coupling constant. 

The discovery of asymptotic freedom~\cite{GrossWilczek,Politzer} in 1974 showed that the coupling strength of non-Abelian gauge theories decreases toward zero at large momenta.  Since deep inelastic electron nucleon scattering experiments carried out in the late 60's indicated just such a behavior called scaling,  the discovery boosted QCD to the leading candidate of the theory of strong interactions.  

A beautiful prediction of asymptotic freedom is the existence of logarithmic violation of scaling which can be quantitatively calculated {\it via} renormalization group methods.  The prediction was later confirmed by experiments, thus establishing the validity of QCD beyond doubt at high energies. 

Since asymptotic freedom at high energies means that the coupling strength increases in the opposite limit of low energies, it was natural to speculate that QCD also provided a solution to the long standing puzzle that quarks had never been observed in experiments.  However, quantum field theory at the time, though quite sophisticated, did not possess means to analyze the behavior of QCD for large coupling constant expected at low energies. 

\subsection{Formulation of QCD on a space-time lattice}
\label{sec:1-2}

An essential ingredient for a strong coupling analysis is a mathematically well-defined formulation of QCD in which ultraviolet divergences  are controlled.  Kenneth Wilson approached this problem with three key ideas: (i) use Euclidean space-time with imaginary time rather than Minkowski space-time with physical time, (ii) replace Euclidean space-time continuum by a discrete 4-dimensional lattice to control ultraviolet divergence, and (iii) maintain local gauge invariance as the guiding principle to construct field variables and action on the lattice.  

The use of Euclidean space-time brings out a beautiful and powerful connection between quantum field theory and statistical mechanics.  This connection was realized in the 60's by Kurt Symanzik and others from the viewpoint of rigorously defining quantum field theory~\cite{Symanzik1969}.  The explosive development in the theory of critical phenomena due to Leo Kadanoff, Michael Fisher, Wilson himself and others in the late 60's and early 70's brought the importance of the concept to the foreground~\cite{WilsonKogut1973}.  A formal proof that quantum field theory in real time can be recovered from that in imaginary time under a set of axioms was given in the first half of 70's~\cite{OsterwalderSchrader1973}.   This connection, then, was a hot topic at the time and Wilson consciously exploited it in his work on lattice QCD. 

\begin{figure}
\begin{center}
    \includegraphics[bb= 0 0 566 218, width=0.9\textwidth]{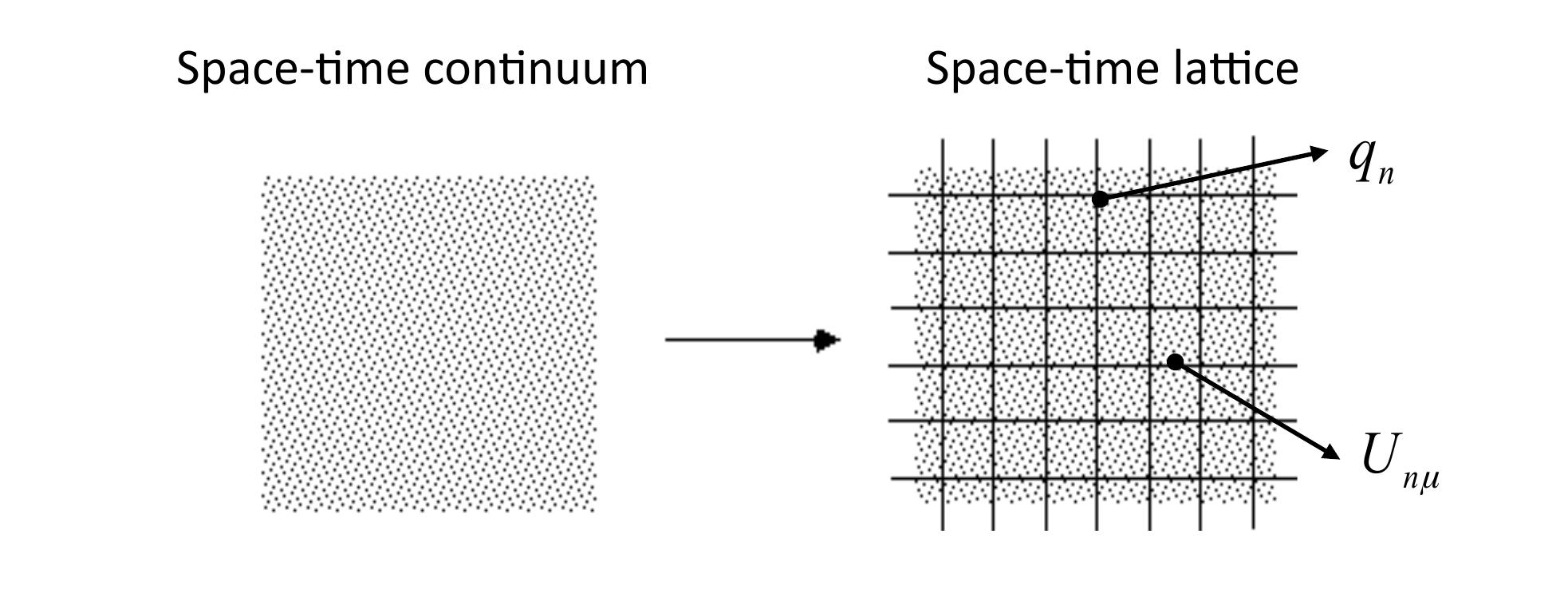}
\end{center}
\caption{Replacing continuum space-time by a space-time lattice. Quark fields are placed at lattice sites, and gluon fields on links.}
\label{fig:lattice}       
\end{figure}

Let us consider a simple cubic 4-dimensional lattice in Euclidean space-time.  The lattice points, called sites, are labeled by an integer component 4-vector $n=(n_1, n_2, n_3, n_4)\in {\bf Z}^4$, and have a physical coordinate $x=na$ with $a$ the lattice spacing.  A pair of neighboring sites at $n$ and $n+\hat{\mu}$ with $\hat{\mu}$ the unit vector in the direction $\mu=1,2,3,4$ are connected by the link between the two sites which may be denoted as $\ell=(n, \mu), n\in{\bf Z}^4, \mu=1, 2, 3, 4$.  An elementary square, or plaquette,  on the lattice is then labelled by $P=(n, \mu, \nu)$ with $n$ the lowest corner and  the two directions spanning the square denoted by $\mu$ and $\nu$.  The lattice construction is illustrated in Fig.~\ref{fig:lattice}.

It is natural to place a quark field $q(x)$ on each site $q(x)=q_n$ with $x=na$.  If one replaces the derivative $\partial _\mu q(x)$ by a finite difference $(q_{n+\hat{\mu}}-q_{n-\hat{\mu}})/2a$,  the quark Lagrangian on a lattice will contain a bilocal term $\overline{q}_nq_{n\pm\hat{\mu}}$.  

In the continuum space-time, bilocal quantities such as $\overline{q}(y) q(x)$ are rendered gauge invariant by inserting a path ordered phase factor $U(y,x)=P\exp(i\int_x^yA_\mu(x)dx_\mu)$, which transforms as $U(y,x)\to V^\dagger(y)U(y,x)V(x)$ under the local gauge transformation. 
Wilson proposed to employ the phase factor along the lattice link $U_{n\mu}\sim P\exp(i\int_{na}^{(n+\hat{\mu})a}A_\mu(x)dx_\mu)$ as the fundamental gluon field variable on the lattice.  
The lattice quark action can then be made invariant by inserting appropriate $U_{n\mu}$'s to the bilocal terms.  One thus finds for the quark action on the lattice, 
\begin{eqnarray}
\label{eq:latticeaction}
S_{\rm quark}= \frac{a^3}{2}\sum_{n\mu}\left(\overline{q}_n\gamma_\mu U_{n\mu}q_{n+\hat{\mu}}-\overline{q}_n\gamma_\mu U_{n-\hat{\mu}\,\mu}^\dagger q_{n-\hat{\mu}}\right)
+ a^4\sum_n\overline{q}_n m_0 q_n,
\end{eqnarray}
with $m_0$ the bare quark mass matrix. 

A classical action for gluons which reduces to the continuum action in the limit $a\to 0$ can be constructed by taking the product of $U_{n\mu}$'s around the boundary of a plaquette $P$~\cite{Wilson1974}:  
\begin{eqnarray}
S_{\rm gluon}&=&\frac{1}{g_0^2}\sum_{P}{\rm Tr}\left[\prod_{n\mu\in\partial P}U_{n\mu}\right], 
\end{eqnarray}
where ${\rm Tr}$ stands for trace over SU(3) indices.  Here $g_0$ is the bare gauge coupling constant at the energy scale of the lattice cutoff $1/a$. 

Lattice QCD as quantum theory is defined by the Feynman path integral.  If $O(U,q, \overline{q})$ is an operator corresponding to some physical quantity, the vacuum expectation value over the quantum average is given by 
\begin{eqnarray}
\label{eq:1}
\left< O(U,q, \overline{q})\right> = \frac{1}{Z_{\rm QCD}}\int\prod_{n\mu}dU_{n\mu}\int\prod_n d\overline{q}_n dq_n O(U,q, \overline{q}) \exp (S_{\rm QCD})
\end{eqnarray}
with 
\begin{eqnarray}
\label{eq:partition}
Z_{\rm QCD}=\int\prod_{n\mu}dU_{n\mu}\int\prod_n d\overline{q}_n dq_n \exp (S_{\rm QCD}),
\end{eqnarray}
where
\begin{eqnarray}
S_{\rm QCD}=S_{\rm gluon}+S_{\rm quark}.
\end{eqnarray}

The gluon link variable $U_{n\mu}$ takes values in the group SU(3) rather than the Lie algebra for the vector gluon field $A_\mu(x)$.  The integration over $U_{n\mu}$'s should be defined as invariant integration over the group SU(3).  Since SU(3) has a finite volume under this integration, the lattice Feynman path integral is well-defined without gauge fixing.  

The integration over the quark fields also requires some care.  Since quarks are fermions, the path integral has to be defined in terms of Grassmann numbers which anticommute under exchange, {\it i.e.,} $q_n q_{m} = - q_{m} q_n$.  These points are clearly spelled out in the Wilson's original paper in 1974~\cite{Wilson1974}. 
 
\subsection{Confinement of color}
\label{sec:1-3}

Whether quarks are confined in QCD can be examined if one knows how to calculate the energy of an isolated quark in interaction with the gluon fields: an isolated quark would exist only if the energy of that state is finite. An elegant method invented by Wilson is to consider a pair of static quark and antiquark, which is created at some point in space-time, then separated to some distance, stays in that configuration for some time, and finally is brought together to a point and annihilated.  Geometrically, the space-time trajectory of the pair forms an oriented closed loop $C$.  Since the color charge of the quark interacts with gluon fields, the creation and annihilation of the pair inserts a phase factor 
\begin{eqnarray}
W\left[C\right] = P\exp\left( i\oint_C dx_\mu A_\mu(x) \right)
\end{eqnarray}   
in the path integral where $P$ indicates ordering along the path.  This is the Wilson loop operator.  

\begin{figure}
  \begin{center}
   \includegraphics[bb= 0 0 704 351,width=0.8\textwidth]{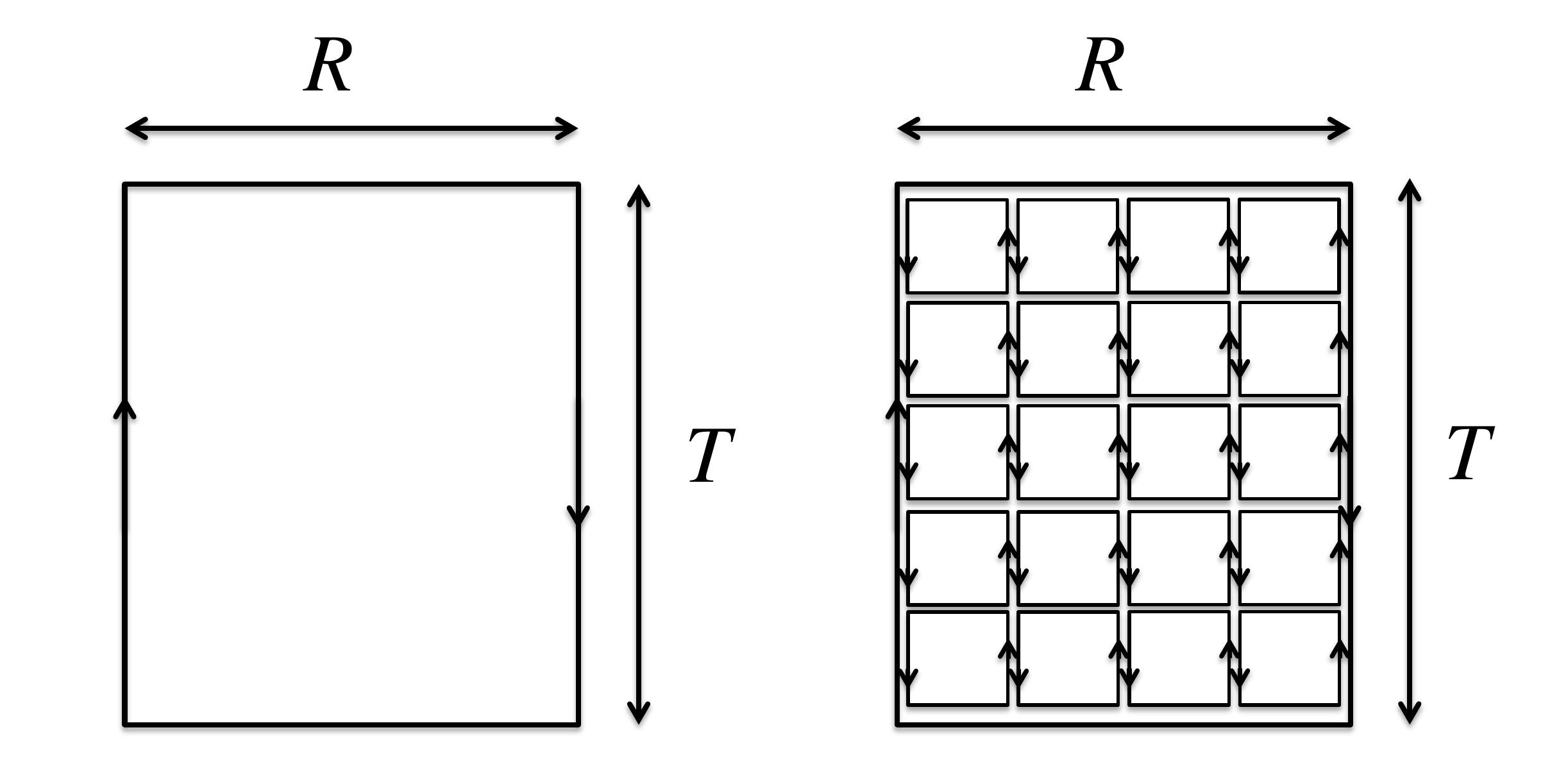}
  \end{center}
  \caption{Left panel: Wilson loop for a temporal size $T$ and a spatial size $R$ measures the potential energy for a pair of static quark and antiquark.  Right panel: The leading order of the strong coupling expansion is obtained by tiling the Wilson loop by plaquettes. }
  \label{fig:wilsonloop}
\end{figure}

If the loop has the shape of a rectangle of a width $R$ in the spatial direction and an extent $T$ in the temporal direction as shown in the left panel of Fig.~\ref{fig:wilsonloop},  the quantum average of the Wilson loop operator measures the energy $E(R)$ of the quak-antiquark pair relative to the vacuum over a temporal length $T$ so that
\begin{eqnarray}
\left< W\left[R\times T\right]\right>\propto \exp\left(-E(R)T\right),\quad T\to\infty.
\end{eqnarray}

Wilson speculated that for large values of the gauge coupling, fluctuations of gluon fields would be large, leading to a significant cancellation among the contributions to the Wilson loop.   This would mean that the Wilson loop average  rapidly decreases for larger loops, or equivalently the probability decreases that the quark and antiquark are found in a well-separated configuration.  This would mean confinement.  Another equivalent statement would be that the energy $E(R)$ grows for larger separations $R$, so that separating quark from antiquark is not possible.  

An amazingly simple calculation suffices to verify this picture if one employs lattice QCD.  Let us assume, following Wilson,  that  quark fields do not play an essential role.  When the bare coupling $g_0$ becomes large, the gluon path integral in (\ref{eq:1}) can be calculated by an expansion in inverse powers of $g_0^2$.  The leading term is obtained when one tiles the $R\times T$ surface of the Wilson loop by a set of plaquettes from the expansion of the gluon part of the weight $\exp\left( S_{\rm gluon}\right)$, as shown in the right panel of Fig~\ref{fig:wilsonloop}.  We then find that
\begin{equation}
\left< W\left[R\times T\right]\right> =N_c\left(\frac{1}{N_cg_0^2}\right)^{RT/a^2}\left(1+O(g_0^{-8})\right), \quad g_0^2\to\infty,
\end{equation}
with $N_c=3$ for SU(3) so that
\begin{eqnarray}
\label{eq:strongcoupling}
E(R)\to \sigma R, \quad R\to\infty \quad {\rm with } \quad \sigma=\frac{1}{a^2}\left(\log\left(N_cg_0^2\right)+O(g_0^{-8})\right),
\end{eqnarray}
namely, a static pair of quark and antiquark are bound by a potential linearly rising with the separation.  Hence they cannot be separated to infinite distance with any finite amount of energy.  

A closed loop $C$ has two geometrical characteristics, the length of the loop $L[C]$ and the area of the minimal surface $A[C]$ spanned by the loop.  The confinement behavior corresponds to an area decay for large loop;
\begin{eqnarray}
\label{eq:area}
\left<W\left[C\right]\right> \propto \exp\left(-\sigma A[C]\right).
\end{eqnarray}
If, on the other hand, the Wilson loop expectation value decays with the loop length,
\begin{eqnarray}
\label{eq:perimeter}
\left<W\left[C\right]\right> \propto \exp\left(-\mu L[C]\right),
\end{eqnarray}
the energy of a quark-antiquark pair saturates to a constant $E(R)\to \mu$ for large separation $R\to\infty$.  Hence there is no longer confinement.  The confining and non-confining phases of gauge theories are thus distinguished by the behavior of the Wilson loop expectation value.    

The possibility that there can be both confining and non-confining phases raises an interesting question whether a confining phase can turn into a non-confining phase if some parameters of the theory are varied.  Temperature is such an important parameter.  As we discuss in more detail in Sec.~\ref{sec:3-4}, the confining property becomes lost through a phase transition when the temperature is raised sufficiently. 

\subsection{Continuum limit}
\label{sec:1-4}

Let us go back to the strong coupling result in (\ref{eq:strongcoupling}).  If one uses the phenomenological value $\sqrt{\sigma}\approx 440 {\rm MeV}$, and if one assumes that a value $\log\left(N_cg_0^2\right)\sim O(1)$ is sufficient for the strong coupling expansion to converge, the corresponding value of the lattice spacing equals $a\sim 0.5$~fm with 1~fm$=10^{-15}$~m.   This value is comparable to a typical length scale of the strong interactions, {\it e.g.,} the charge radius of proton $R_p\sim 0.9$~fm.  We certainly know that space-time is continuous well below such length scales.   Thus one need to know if confinement holds for smaller lattice spacings.  

\begin{figure}
  \begin{center}
   \includegraphics[bb= 0 0 228 239,width=0.5\textwidth]{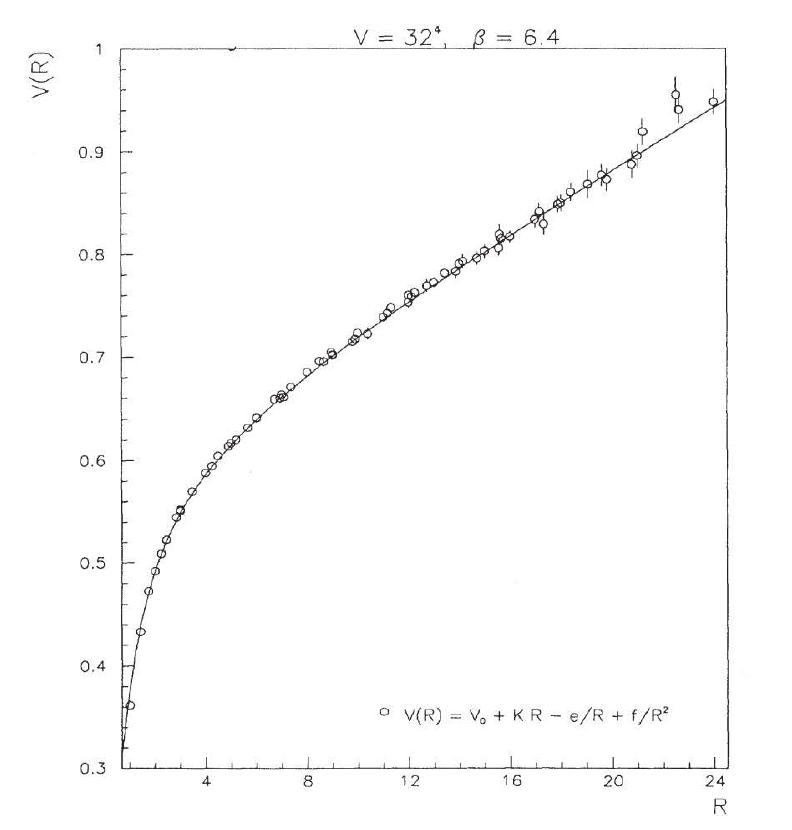}
  \end{center}
  \caption{Monte Carlo result for the static potential energy $E(R)$ calculated in pure gluon theory showing a linearly rising term at long distances and a Coulomb term at short distances. From Ref.~\protect{\cite{BaliShilling1993}}. }
  \label{fig:potential}
\end{figure}

In Fig.~\ref{fig:potential} we show the static potential $E(R)$ calculated by Monte Carlo simulation of the pure gluon theory~\cite{BaliShilling1993}.  We observe a linearly rising potential $E(R)\sim \sigma R$ at large distances.  There is also a Coulomb behavior $V(R)\sim \alpha/R$ at short distances, which is consistent with a perturbative one gluon exchange expected from asymptotic freedom.  The lattice spacing estimated from  the phenomenological value $\sqrt{\sigma}\approx 440$~MeV is $a\approx 0.0544(4)$~fm.    This is one of many evidences that the property of confinement holds not just at strong coupling but also toward weak couplings with small lattice spacings. 

The question then is whether the confinement property really persists as the lattice spacing is taken toward the limit of continuous space-time $a\to 0$.  For Wilson, who elucidated critical phenomena with his renormalization group ideas, this was not a conceptually difficult issue.   The strong coupling calculation demonstrates that the system of gluons without quarks is in a confining phase.  When one decreases the coupling constant $g_0$, this property persists as long as one does not encounter a phase transition to a non-confining phase characterized by perimeter decay (\ref{eq:perimeter}).  Suppose that there is such a second-order phase transition at $g_0^2=g_c^2$.  Close to the critical point, the correlation length measured in lattice units diverges $\xi(g_0)/a\to\infty$ as $g_0\to g_c$.  This means that one should be able to hold the physical correlation length $\xi(g_0)$ fixed to a physical value observed in experiment while sending the lattice spacing to zero $a\to 0$, recovering physics in the space-time continuum.  

One can define lattice gauge theory for a variety of groups and space-time dimensions.  As with the case for spin systems, the existence and position of second-order phase transition points would depend on them.  For the group SU(3) in 4 dimensions, a particularly attractive possibility is $g_c=0$. In fact, this is the only possibility if confinement at low energies and asymptotic freedom at high energies are to coexist in the same phase.  

The renormalization group allows a more concrete discussion.  The change of gauge coupling $g_0\to g_0+dg_0$ under a change of cutoff scale $a\to a+da$ while fixing a physical scale constant defines a renormalization group $\beta$ function 
\begin{eqnarray}
\label{eq:beta}
a\frac{dg_0(a)}{da} = -  \beta(g_0),
\end{eqnarray}
where the minus sign is inserted to be compatible with the conventional definition using momentum slicing.  The strong coupling result (\ref{eq:strongcoupling}) shows that $\beta(g_0)$ is large and negative for large $g_0$.  For small values of $g_0$, one can employ perturbation theory to confirm the asymptotic freedom result,
\begin{eqnarray}
\label{eq:two-loop}
\beta(g_0)=-b_0g_0^3-b_1g_0^5+O(g_0^7)
\end{eqnarray}
with  
\begin{eqnarray}
b_0&=&\frac{1}{(4\pi)^2}\left(\frac{11}{3}N_c-\frac{2}{3}N_f\right),\\ 
b_1&=&\frac{1}{(4\pi)^4}\left(\frac{34}{3}N_c^2-\frac{N_c^2-1}{N_c}N_f-\frac{10}{3}N_cN_f\right),
\end{eqnarray}
for $SU(N_c)$ gauge group and $N_f$ flavors of fermions in the fundamental representation. The first two coefficients are negative  for our world with $N_c=3$ and $N_f=6$.  A natural supposition is that the beta function is negative for all values of the coupling, with the only zero residing at $g_c=0$.  

Wilson  started a numerical Monte Carlo calculation to check this by a block spin renormalization group for gauge group SU(2)~\cite{Wilson1979}.  This attempt was followed by several serious calculations with SU(3) gauge group in the 80's. The results, though mostly restricted to the pure gluon theory and had fairly large errors,  showed that the beta function stayed negative and became consistent with the two-loop result (\ref{eq:two-loop}) toward weak coupling (see, {\it e.g.},  Refs.~\cite{Bowleretal1986,Guptaetal1988}).  

\begin{figure}
  \begin{center}
   \includegraphics[bb= 0 0 294 254,width=0.5\textwidth]{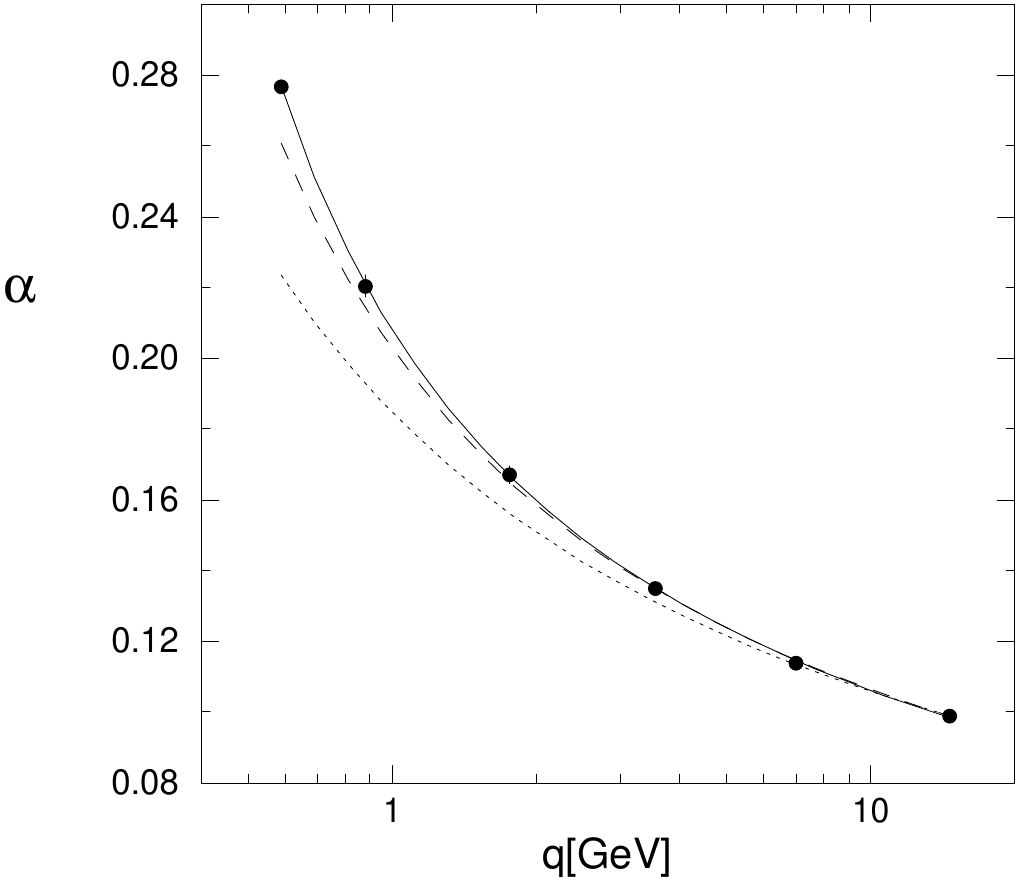}
  \end{center}
  \caption{Running of renormalized coupling constant $\alpha(q^2)=g^2(q^2)/4\pi$ in the Schr\"odinger functional scheme as a function of momenta scale $q$.  Dashed and dotted lines are two- and one-loop evolution starting from the right-most point.  The physical scale is fixed by the Sommer scale $r_0=0.5$~fm.  From Ref.~\protect{\cite{Luscheretal1994}}. }
  \label{fig:running}
\end{figure}

A different, and perhaps a more elegant,  approach was developed in the 90's by the Alpha Collaboration~\cite{Luscheretal1994}.  Their method of step scaling function defines the renormalized coupling constant at a scale $1/L$ through the Schr\"odinger functional on a lattice of a finite size $L^3\times T$ with a prescribed boundary condition in time, and follows the evolution of the coupling under a change of scale by a factor 2.  The continuum limit is systematically taken in this process.  Therefore, the end result is the evolution of the renormalized coupling from low to high energies in the continuum theory.  

In Fig.~\ref{fig:running} we show the result for the pure gluon theory (solid circles)  and a comparison with perturbation theory (dashed and dotted lines).  The full evolution runs from 10~GeV down to a few hundred MeV, where the system is in a strong coupling non-perturbative regime.  Thus the confining behavior at large distance is continuously connected with the asymptotically free behavior at short distances.  The full evolution runs almost parallel to the two-loop evolution.  Similar results have been obtained in full QCD with $N_f=2, 3, 4$ dynamical quarks~\cite{ALpha2005,PACS-CS2009a,AlphaNf4}.

Based on the results described above, one can  state, though a mathematically rigorous proof is yet lacking, that QCD is in the confining phase over the entire range of coupling from zero to infinity.  

\subsection{Quarks and chiral symmetry}
\label{sec:1-5}

\subsubsection{Chiral symmetry}
\label{sec:1-5-1}

An important feature of the strong interaction which was recognized in the late 50's and early 60's is chiral symmetry.  Studies in this period led to the concept of spontaneous breakdown of symmetry, and  a realization that it is accompanied by the emergence of massless bosons, called Nambu-Goldstone bosons today.  

In QCD language, chiral symmetry is invariance under the global transformation $q(x)\to Wq(x)$, where $W=\exp( i\alpha\gamma_5) \in {\rm SU(N_f)}$ acts on the Dirac-flavor indices and rotates left and right handed chiral components of $q(x)$ in the opposite direction.  This  symmetry is explicitly broken by the quark mass term.  Hence it holds approximately for three light quarks, up, down and strange.   It is not very relevant for the heavier quarks, charm, bottom and top.   The octet of pseudo scalar mesons, $\pi, K, \eta$, are identified as the Nambu-Goldstone bosons corresponding to spontaneous breakdown of chiral symmetry. With a strongly interacting  dynamics  at large distances, QCD should dynamically explain spontaneously broken chiral symmetry and its physical consequences.  

\subsubsection{chiral symmetry on a space-time lattice}
\label{sec:1-5-2}

Introduction of a space-time lattice affects bosons and fermions in different ways.  This is most easily seen by looking at the kinetic term in the free field case.  For a boson field $\phi_n$, the second-order derivative $-\partial_\mu^2\phi(x) $ is discretized as  $-(\phi_{n+\hat{\mu}}+\phi_{n-\hat{\mu}}-2\phi_n)/2a^2$, whereas for a fermion field $q(x)$, the first order derivative $\gamma_\mu\partial_\mu q(x)$ is discretized as $\gamma_\mu(q_{n+\hat{\mu}}-q_{n-\hat{\mu}})/2a$.  The momentum space expression then becomes $2(1-\cos(p_\mu a))/a^2$ and $i\gamma_\mu\sin(p_\mu a)$, respectively.  There is a crucial difference in the location of the zeros in the Brillouin zone: the boson case has a zero only at $p_\mu=0$ for all $\mu$ whereas the fermion case has a set of zeros for $p_\mu=0$ or $\pi/a$ for each $\mu$.  Since the zeros gives rise to poles in the propagator, a naive fermion discretization leads to  multiple copies of the state at $p_\mu=0$.  This is the fermion species doubling problem. 

Building upon a pioneering work by Karsten and Smit~\cite{KarstenSmit1981}, Nielesen and Ninomiya~\cite{NielsenNinomiya1981} proved that essentially the same conclusion holds under a set of rather general conditions.  The theorem states that if the fermion action satisfies (i) chiral symmetry $q_n\to\exp(i\alpha\gamma_5)q_n$, (ii) invariance under unit translation on the lattice, (iii) Hermiticity, and (iv) locality, then the spectrum of fermions contains even number of particles, half of them left handed and the other half right handed.  

An elegant  topological proof runs as follows~\cite{karsten}.  If we write a Dirac fermion action on a lattice in a general form, $S_F=\sum_{n,m}\overline{q}_nD_{n,m}q_m$, the assumptions (i) and (ii) imply that $D_{n,m}=\sum_{\mu}\gamma_\mu F_\mu(n-m)$ where  $F_\mu(n)$ is a vector function which, by (iii)  satisfies $F_\mu^\dagger(n)=F_\mu(-n)$, and by (iv) rapidly decreases as $\vert n\vert$ becomes large.  The Fourier transform $\tilde{F}_\mu(p)$ is, therefore, a well-defined real vector field over the Brillouin zone in momentum space.  Let $p=p^{(i)}, i=1,2,\cdots$ be the zeros of the vector field $\tilde{F}_\mu(p)=f_{\mu\nu}(p_\nu-p^{(i)}_{\nu})+O((p-p^{(i)})^2)$.    They correspond to the poles of the propagator $D^{-1}(p)$, and hence to the particle states.  The relative chirality of these states are determined by the index,  ${\rm sign}( \det f_{\mu\nu})$.  Since the sum of index of a vector field over the 4-dimension torus vanishes by the Poincar\'e-Hopf theorem, there has to be an equal number of fermion states with opposite chirality.  
 
The Nielsen-Ninomiya theorem  indicates that  one either has to abandon chiral symmetry or one has to allow for the presence of species doubling.  Wilson's choice, which he actually wrote down~\cite{Wilson1975} prior to the publication of the theorem, was to add a term which softly breaks the chiral symmetry of the naive lattice action in (\ref{eq:latticeaction}):
\begin{eqnarray}
\label{eq:wilson}
\delta S_{\rm W}&=&\frac{a^3}{2}\sum_{n\mu}\left(2\overline{q}_nq_n-\overline{q}_nU_{n\mu}q_{n+\hat{\mu}}-\overline{q}_nU_{n-\hat{\mu}, \mu}^\dagger q_{n-\hat{\mu}}\right).
\end{eqnarray}
For the free field case, this term adds $\sum_\mu (1-\cos p_\mu a)/a$ to the kinetic term $i\sum_\mu \gamma_\mu\sin p_\mu a$, and hence removes the zeros at $p_\mu=\pi /a$.  

Let us add that toward the continuum limit $a\to 0$ the Wilson's added term becomes of form $a\overline{q}(x)D_\mu^2q(x)$ relative to the original term $\overline{q}(x)i\gamma_\mu D_\mu q(x)$ in (\ref{eq:latticeaction}), {\it i.e.}, higher order in $a$.  Apart from removing the doublers, the effect of the added term disappears in the continuum limit. 

Chiral breaking effects of the Wilson's added term can be analyzed by the method of Ward identities~\cite{Bochicchioetal1985}.  In particular, a definition of quark mass can be given that satisfies the PCAC relation~\cite{Bochicchioetal1985,ItohIwasakiOyanagiYoshie1986}.   A detailed  analysis of the phase structure on the $(g_0, m_0)$ plane was made~\cite{Kawamoto1981},  and the existence of a massless pion, in spite of chiral symmetry breaking, was explained as due to spontaneous breakdown of parity Z(2) symmetry~\cite{Aoki1984}.  With these analytical developments, Wilson's formulation provides a quantitative framework for the computation of physical observables, and is used extensively in Monte Carlo studies.   

A variant of Wilson's formulation is to consider two flavors of quarks as a pair and add a twisted mass term to the naive action (\ref{eq:latticeaction})~\cite{tmQCD2001}:
\begin{eqnarray}
\delta S_{\rm tm}=a^4\sum_n\overline{q}_n i\mu_q\gamma_5\tau_3 q_n,
\end{eqnarray}
where $\tau_3$ acts on the flavor index.  This formulation has an attractive feature that one can twist the angle $\alpha=\arctan \mu_q/m_0$ in such a way that $O(a)$ lattice artifacts are absent in physical observables~\cite{tmQCD2004,AokiBaer2006}.  Large-scale simulations are being made using such "maximally" twisted QCD. 

There is a different  method, called the staggered formulation~\cite{Susskind1977},  which retains a U(1) chiral symmetry at the cost of a four fold species doubling.  In the 4-dimensional Euclidean formulation~\cite{SharatchandraThusWeisz1981}, it starts with a single component fermion field $\chi_n$ at each site, and reconstructs four species of Dirac fields $\psi =(\psi_1, \cdots, \psi_4)$ from 16 $\chi$'s on the 16 vertices of a 4-dimensional unit hypercube~\cite{KlubergSternetal1981}.  The original action is invariant under an even-odd U(1) symmetry $\chi_n\to\exp(i(-1)^{\vert n\vert}\alpha)\chi_n$, which translates into an axial U(1) transformation on the four Dirac fields of form $\psi\to\exp(i\alpha\gamma_5\otimes\xi_5)\psi$ where $\xi_5$ acts on the species index $\alpha$ of the Dirac field $\psi_\alpha, \, \alpha=1, \cdots, 4$.   

It is generally believed that lattice QCD with the staggered fermion formalism converges to continuum QCD with $N_f=4$ degenerate flavors of quarks~\cite{Sharpe2006}.  An elaborate effective theory description has been developed to control the breaking of the full 4 "flavor" chiral symmetry down to the U(1) subgroup at finite lattice spacing~\cite{LeeSharpe1999}.  On these theoretical bases, the staggered formalism is also extensively used in Monte Carlo simulations.  

\subsubsection{Lattice fermion action with chiral symmetry}
\label{sec:1-5-3}

In 1981, a year after Nielsen and Ninomiya presented their theorem, Wilson revisited the issue of chiral symmetry from a different perspective with Ginsparg~\cite{Wilson1981}.  He asked what would be the relation satisfied by a lattice Dirac operator $D_{n,m}$ if it were derived by a (chiral symmetry breaking) block spin transformation from a chiral invariant theory.  The answer turned out to be remarkably simple; it is given by 
\begin{eqnarray}
\label{eq:GW}
\gamma_5 D_{n,m}+D_{n,m}\gamma_5 = a \sum_k D_{n,k}\gamma_5 D_{k,m}.
\end{eqnarray}
Since $D$ does not anticommute with $\gamma_5$,  assumption (i) of the Nielsen-Ninomiya theorem is not satisfied, and so the conclusion of the theorem does not hold.  In fact, there is no species doubling.  Furthermore,  the axial vector current for this action has the correct U(1) anomaly.   

One can rewrite (\ref{eq:GW}) in terms of the propagator $G=D^{-1}$ as 
\begin{eqnarray}
\label{eq:GWI}
\gamma_5 G_{n,m}+G_{n,m}\gamma_5 = a\delta_{nm}\gamma_5.
\end{eqnarray}
Hence the breaking of chiral symmetry is a local effect.  One may expect then that a modified form of symmetry may exist.  This was found almost 20 years later~\cite{Luescher1998}.  Fermion actions that satisfy the Ginzparg-Wilson relation (\ref{eq:GW}) are invariant under an infinitesimal transformation given by 
\begin{eqnarray}
\delta\overline{\psi}=\overline{\psi}(1-\frac{1}{2}aD)\gamma_5,\quad \delta\psi=\gamma_5 (1-\frac{1}{2}aD)\psi.
\end{eqnarray}

Several forms of fermion action which satisfy the Ginsparg-Wilson relation were discovered in the 90's. 
One form is a domain-wall formalism~\cite{Kaplan1992,FurmanShamir1995} in which the 4-dimensional fermion field is constructed as the zero mode of a 5-dimensional theory generated by a mass defect at the boundary in a fictitious fifth dimension. 
Another form is given by the overlap formalism~\cite{NarayananNeuberger1995,Neuberger1998}.  In this case an explicit form of the operator $D$ is given by 
\begin{eqnarray}
D=\frac{1}{a}\left(1+\gamma_5H(H^\dagger H)^{-1/2}\right), \quad H=\gamma_5 (aD_W-1)
\end{eqnarray}
with $aD_W+m_0$ the operator for the Wilson fermion action  with a negative mass $m_0=-1$.   The two forms are equivalent in the limit of infinite fifth dimension~\cite{Neuberger1998-2,Borici1999}.  

Yet another form is the perfect action~\cite{HasenfratzNiedermeyer1998}, so named because it is  defined as the fixed point of a block spin transformation of renormalization group for QCD.   This form follows from the line of reasoning of Ginsparg and Wilson, but it was pursued and arrived at independently almost two decades later.  

All forms of action, particularly the domain wall and overlap actions,  have come to be used extensively in the last decade.  The domain wall formalism has been exploited by RBC-UKQCD Collaborations (see, {\it e.g.,} \cite{RBCUKQCDDomainwall2013}), and the overlap formalism by JLQCD (see \cite{KEKOverlap2012} for a recent review).    The situation with the perfect action is reported in \cite{Hasenfratzetal2005}. 

\subsubsection{Spontaneous breakdown of chiral symmetry}
\label{sec:1-5-4}

Spontaneous breakdown of chiral symmetry is best studied by examining the behavior of the order parameter of chiral symmetry.  This is given by the quark bilinear operator $\Sigma=\left<\overline{q}_nq_n\right>$.  If $\Sigma\ne 0$ after sending the spatial volume to infinity $V\to\infty$ followed by the limit of quark mass to zero $m_q\to 0$, then chiral symmetry is spontaneously  broken.  

\begin{figure}[t]
  \begin{center}
   \includegraphics[bb= 0 0 360 252,width=0.6\textwidth]{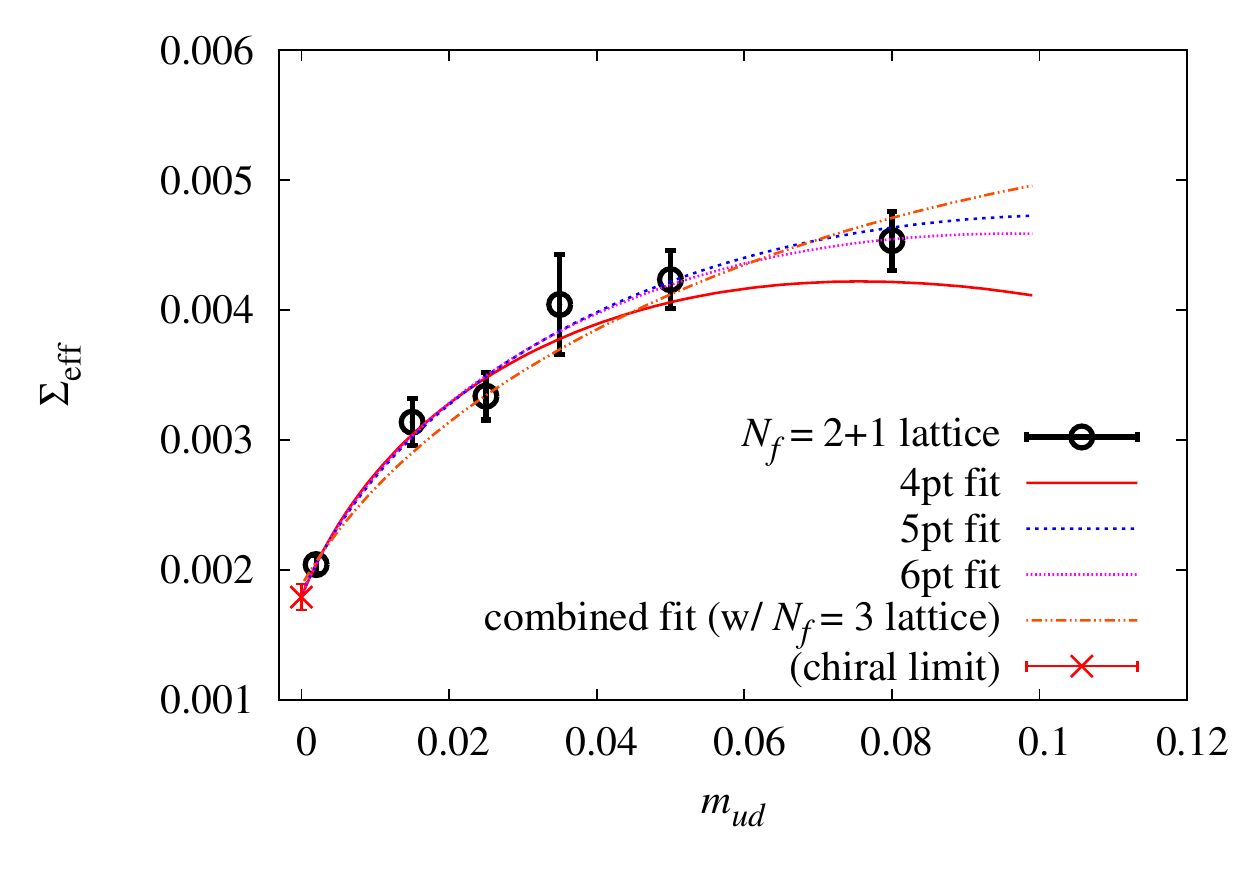}
  \end{center}
  \caption{Chiral order parameter $\Sigma=\left<\overline{q}_nq_n\right>$ as a function of degenerate up-down quark mass in lattice units in $N_f=2+1$ flavor lattice QCD with overlap fermion formalism.  The strange quark mass is held fixed at a value close to experiment.  From Ref.~\protect{\cite{KEKOverlap2012}}.  Color online. }
  \label{fig:chiral}
\end{figure}

In Fig.~\ref{fig:chiral} we show the result for $\Sigma$ as a function of the degenerate up-down quark mass $m_{ud}$ in lattice units obtained with the overlap formulation for $N_f=2+1$ lattice QCD.  We choose these data since the overlap formalism is the cleanest regarding the chiral aspect among lattice fermion formulations.  The terminology $N_f=2+1$ refers to the fact that the up and down quark masses are taken degenerate, while the strange quark mass has a separate value.  Strictly speaking, the values shown in Fig.~\ref{fig:chiral} are obtained from the eigenvalue distribution of the Dirac operator and not by calculating the condensate directly.   

The concave curvature as a function of  $m_{ud}$ is consistent with the presence of a logarithm term predicted by chiral perturbation theory.  Extrapolating to $m_{ud}=0$ including the effect of the logarithm yields a non-zero value, supporting spontaneous breakdown of chiral symmetry.  The dependence of the pion mass is consistent with $m_\pi^2\propto m_{ud}$ (up to logarithmic corrections), as follows from the Nambu-Goldstone theorem.    

Similar results have been obtained for the staggered fermion formalism which has U(1) chiral symmetry.  The analysis is more complicated for the Wilson fermion formulation, since one needs to carry out a subtractive renormalization for both chiral condensate and quark mass due to the soft chiral symmetry  breaking induced by the Wilson term (\ref{eq:wilson}).  Once these subtractions are done, one finds the signatures expected for spontaneous breakdown of chiral symmetry, such as the relation $m_\pi^2\propto m_{ud}$ for an appropriately defined quark mass~\cite{Bochicchioetal1985,ItohIwasakiOyanagiYoshie1986}.

\subsection{Heavy quarks on a lattice}
\label{sec:1-5-5}

The three light quark quantum numbers, {\it i.e.},  up, down and strange, have been known from 1930's and 1950's.  In contrast,  heavier quarks were discovered more recently, {\it i.e.,} charm quark in 1974, bottom quark in 1977, and top quark in 1995.  These heavy quarks occupied the central place in the experimental and theoretical studies toward establishing the Standard Model, particularly with the construction of B factories in the 90's and experiments with them in the 2000's. 
 
Studying heavy quarks with lattice QCD poses the problem that for a large value of a heavy quark mass $m_h$, the dimensionless combination $m_ha$ is also large, leading to an amplification of lattice discretization errors, especially if $m_ha\gg 1$.  This situation applies to the bottom quark with a mass $m_b\approx 5$~GeV, since lattice QCD simulations to date have been made with inverse lattice spacings in the range  $a^{-1}\approx 2-4$~GeV.  

Several methods have been formulated and employed to deal with this problem.  The static approximation~\cite{Eichten88} is an expansion in $1/m_h$,  NRQCD~\cite{LepageThacker88} is a reformulation of QCD for non-relativistic quarks with an expansion in powers of the quark velocity $v$, and a relativistic formalism~\cite{ElKahdra1997,AokiKuramashi2003,ChristLin2007} modifies the Wilson quark action so as to systematically reduce the effects of large $m_ha$.  

All three methods have been extensively used to calculate physical quantities involving charm or bottom quark.  In particular, matrix elements such as the pseudoscalar decay constants and form factors calculated with these methods have been playing an important role in constraining the Cabibbo-Kobayashi-Maskawa matrix elements including the CP violation phase.  

It should be mentioned that with increasingly smaller lattice spacings becoming accessible with progress of algorithms and computer power, direct simulations are replacing calculations with effective heavy quark theories.  This has already occurred for charm quark, and it may not be too far into the future that bottom quark becomes treated in a similar way. 

\section{Lattice QCD as computation}
\label{sec:2}

\subsection{Numerical simulation and lattice QCD}
\label{sec:2-1}

Lattice QCD offered a framework for conceptually understanding the dynamics of non-Abelian gauge fields.  In particular, it elucidated the mechanism of confinement in a way which would have been impossible in the perturbative framework of field theory.  Nonetheless, calculation methods to obtain physical results did not really exist; for example, higher order strong coupling expansions were very cumbersome and hard to extrapolate to the continuum limit expected at $g_0\to 0$.  

Monte Carlo simulations offered a new approach to solve this impasse.  Of course Monte Carlo  methods had been known since the pioneering era of electronic computers in the late 40' and early 50's.  The Metropolis algorithm to handle multi-dimensional integrations for statistical mechanical systems was formulated in 1953~\cite{Metropolis}.  Applications to spin systems in statistical mechanics started to appear in the 60's and was pursued increasingly in the 70's.  

It is in this context that Creutz, Jacobs and Rebbi carried out a Monte Carlo study of Ising gauge theory with Z(2) gauge group in 4 dimensions in 1979, finding a first-order phase transition~\cite{CreutzJacobsRebbi1979}.  Creutz extended the application to SU(2) gauge group~\cite{Creutz1979}, and extracted the string tension $\sigma$ in front of the area decay of the Wilson loop.  The dependence of $\sigma$ on the gauge coupling $g_0$ turned out to be consistent with the scaling law predicted by the renormalization group.  This suggested the possibility that a continuum limit could be successfully taken, giving rise to a hope that the confinement problem could be solved in a  numerical way. 

Wilson's interest in numerical analyses and computer applications started early in his career~\cite{Wilson2005}.  He often used numerical methods to carry out his analyses.\footnote{Probably his most famous numerical work is a renormalization group solution of the Kondo problem~\cite{Wilson1975-2}.   The numerical rigor he maintained for this work has become legendary. For the universal ratio of the two temperatures $T_K$ and $T_0$  characterizing the high and low temperature scales, he obtained $W/(4\pi)=(4\pi)^{-1}T_K/T_0=0.1032(5)$.  Six years later, an exact solution by the Bethe ansatz yielded $W/(4\pi)=0.102676\cdots$~\cite{AndreiLowenstein1981}, verifying the Wilson's number to the fourth digit within the estimated error!}   In his 1979 lecture at the Carg\'ese Summer School~\cite{Wilson1979}, Wilson reported a  block-spin renormalization group analysis for SU(2) gauge group using Monte Carlo methods to evaluate the necessary path integral averages.   

These attempts introduced a method hitherto unknown in field theory.  The method looked very promising, and immediately attracted attention of particle theorists.  In particular, it was applied to calculate masses of hadrons directly from quarks and gluons.  The method was based on the observation that if $O_{\rm H}(t)$ is an operator at a time slice $t$ corresponding to a hadron H, {\it e.g., }  $\pi^+\sim \overline{u}\gamma_5 d$ for pion or $p\sim ({}^tuC\gamma_5d)d $ for proton, then the 2-point Green's function for this operator behaves  for large times as 
\begin{eqnarray}
\left<O_{\rm H}(t)O_{\rm H}^\dagger(0)\right>\to Z\exp\left(-m_{\rm H}a t\right), \quad t\to\infty, 
\end{eqnarray}
where $m_{\rm H}$ denotes the mass of hadron H.   
Therefore, calculating the two-point function by a Monte Carlo simulation and extracting the slope of the exponential decay would yield the mass  of hadron H.   

This program was carried out by Weingarten~\cite{Weingarten1982}, and by  Hamber and  Parisi~\cite{HamberParisi1981} in 1981.  The lattice size employed was  $12^4$ for the Icosahedral subgroup of SU(2) for the former, and  $6^3\times 12$ for SU(3) for the latter work.   Albeit very modest in today's standards, their studies clearly demonstrated the feasibility of their approach, which accelerated an explosive development of Monte Carlo simulations in lattice QCD. 

\subsection{Massive parallelism and lattice QCD}
\label{sec:2-2}

Monte Carlo simulation of lattice QCD is based on the method of importance sampling.  Let $\phi_n$ be a field variable at a site $n$ on a lattice $\Lambda$ with $V$ lattice points, and $O(\phi)$ an operator.  In field theory, one wishes to calculate the average value of $O(\phi)$ weighted with the action $S[\phi]$ according to, 
\begin{eqnarray}
\left< O(\phi)\right> = \frac{1}{Z}\int\prod_{n}d\phi_n O(\phi) \exp\left(S[\phi]\right),
\end{eqnarray}
where 
\begin{eqnarray}
Z=\int\prod_{n}d\phi_n O(\phi) \exp\left(S[\phi]\right).
\end{eqnarray}

Let us define a configuration $C$ as a point $C=(\phi_1, \cdots, \phi_V)$ in the integration space of all  fields on the lattice $\Lambda$.  Monte Carlo evaluation of the integral proceeds by setting up a stochastic chain of configurations $C_1\to C_2\to\cdots\to C_N $  such that the distribution of configurations converges to the weight of the integral:
\begin{eqnarray}
\rho_N(C)=\frac{1}{N}\sum_{i=1}^N\delta(C-C_i)\to \frac{\exp\left(S[C]\right)}{Z}, \quad N\to\infty.
\end{eqnarray}
In the  well-known Metropolis algorithm, the convergence is guaranteed by accepting or rejecting a new trial configuration $C_ {\rm trial}$ according to the probability $P_{\rm acc}={\rm Min}\left(1, \exp(S[C_{\rm trial}]-S[C_{\rm old}])\right)$ where $C_{\rm old}$ is the latest configuration of the chain. 

A very important property in Monte Carlo calculations of field theoretical systems is locality.  A field variable $\phi_n$ at a site $n$ interacts only with those variables $\phi_m$'s in a limited neighborhood of the site $n$.  In creating a trial configuration $C_{\rm trial}$ from an old configuration $C_{\rm old}$,  and in  calculating the difference of the action $S[C_{\rm trial}]-S[C_{\rm old}]$, the numerical operations at a site $n$ can be carried out independently of those at a site $m$ unless the pair is within the limited neighborhood of each other.  

\begin{figure}
  \begin{center}
   \includegraphics[bb= 0 0 550 210,width=0.9\textwidth]{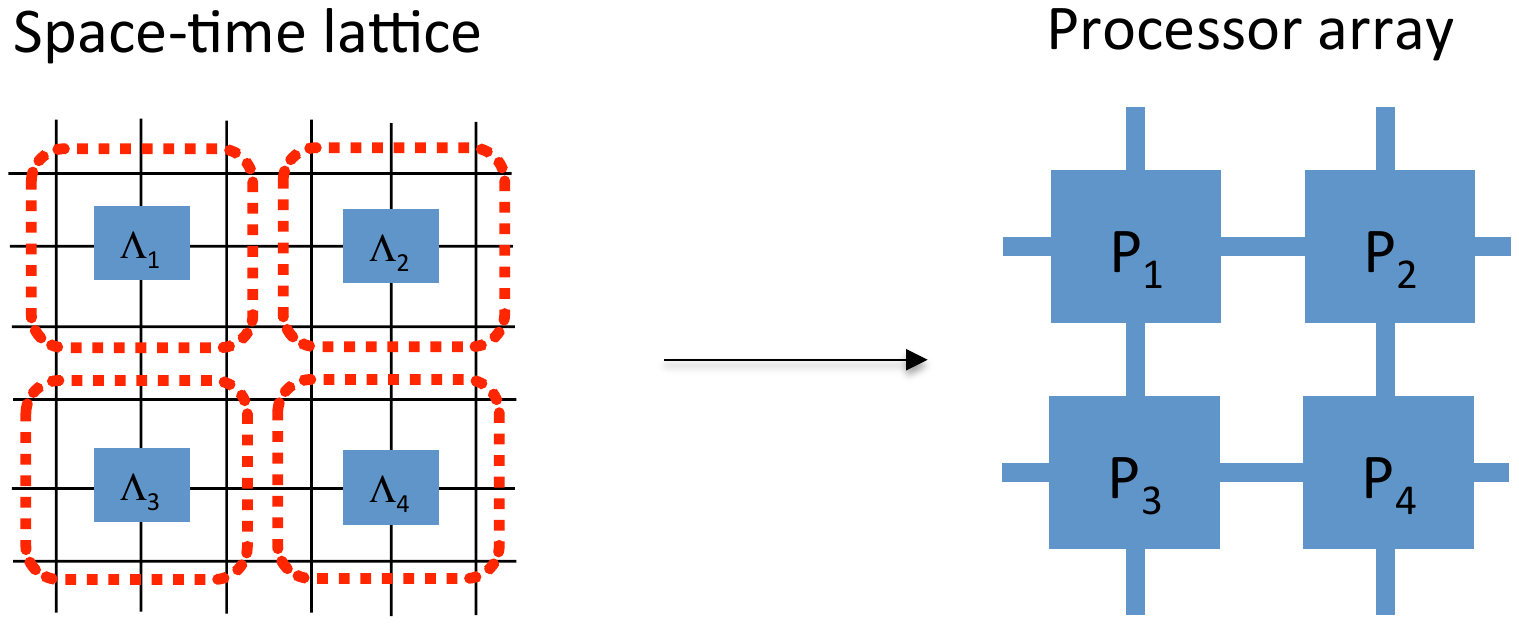}
  \end{center}
  \caption{Mapping of space-time lattice $\Lambda=\cup_{\alpha=1}^{N_P}\Lambda_\alpha$ to an array of processors $P_\alpha, \alpha=1, \cdots, N_P$ interconnected according to the space-time lattice topology.   Color online. }
  \label{fig:lattice-PU}
\end{figure}

As shown in Fig.~\ref{fig:lattice-PU}, let us divide the entire lattice $\Lambda$ into a set of sub-lattices $\Lambda_\alpha, \alpha=1,2,\cdots, N_P$,  and assign each sub-lattice $\Lambda_\alpha$ to a processor $P_\alpha$.  Because of locality, calculations by  the processor $P_\alpha$ can be carried out independently of those by the other processors, except that the processors with overlapping boundaries have to exchange values of $\phi_k$'s in the boundaries before and/or after the calculations in each sub lattice.  This means that, for a fixed lattice size, the computation  time can be reduced by a factor $N_P$, and for a fixed sub-lattice size, one can enlarge the total lattice size proportionately to the number of processors $N_P$ without increasing the computation time.  

This is an ideal case of the parallel computation paradigm.  The locality property, which is one  of the fundamental premises of field theoretic description of our universe, allows a mapping of calculations on a space-time lattice to a parallel array of processors  interconnected with each other according to the connection of space-time sub-lattices.   

\subsection{Parallel computers for lattice QCD}
\label{sec:2-3}

Immediately after Monte Carlo calculations started in lattice QCD, several groups started to plan the building of a parallel computer for lattice QCD calculations.   A crucial factor which helped push such an activity was a rapid development of microprocessors in the 70's.  As shown in Table~\ref{tab:microprocessors}, starting with a 4 bit Intel 4004 in 1971, increasingly more powerful microprocessors were developed and were put out on the market at prices affordable by academic scientific projects.   

In Table~\ref{tab:80s} we list the parallel computers developed by physicists for lattice QCD in the 80's.  Typically, these machines employed a commercial microprocessor such as those listed in Table~\ref{tab:microprocessors} as the control processor, and combined it with a floating point unit to enhance  numerical computation capabilities.  

\begin{table}[b]
\caption{Representative microprocessors developed in the 70's.  Price/chip is from Wikipedia.}
\label{tab:microprocessors}  
\begin{center}
\begin{tabular}{llll}
\hline\noalign{\smallskip}
year & name & bit & price/chip  \\
\noalign{\smallskip}\hline\noalign{\smallskip}
1971 & Intel 4004 & 4 bit & \$ 60 \\
1972 & Intel 8008 &	8 bit &\$120 \\
1974 & Intel 8080 &	8 bit	&\$360         \\ 
1974 & Motorola 6800 &	8 bit &\$360 \\
1978	 & Intel 8086 & 	16 bit & \$320  \\
1979 & Motorola 68000 &32 bit &\\
\noalign{\smallskip}\hline
\end{tabular}
\end{center}
\end{table}

\begin{table}[b]
\caption{Parallel computers dedicated to lattice QCD developed in the 80's. ''year" marks the completion date.  }
\label{tab:80s}  
\begin{center}
\begin{tabular}{llllll}
\hline\noalign{\smallskip}
name & year & authors  & CPU & FPU & peak speed  \\
\noalign{\smallskip}\hline\noalign{\smallskip}
PAX-32*	& 1980&	Hoshino-Kawai &   M6800 &     AM9511	&	0.5 MFlops\\
Columbia	& 1984&	Christ-Terrano &    PDP11 &	TRW &    --  \\
Columbia-16 &	1985 &	Christ et al	&     Intel 80286  &  TRW	&	0.25 GFlops \\
APE1 &	1988 &	Cabibbo-Parisi &     3081/E &	     Weitek &  		1 GFlops \\
Columbia-64 & 	1987 & 	Christ et al &      Intel 80286 &   Weitek &		1 GFlops \\
Columbia-256	&1989&	Christ et al&     M68020	&     Weitek	&16 GFlops\\	
ACPMAPS	&1991&	Mackenzie et al&	micro VAX & Weitek	&5 Gflops\\ 
QCDPAX & 1991 & 	Iwasaki-Hoshino & M68020 &	     LSI-logic &14 GFlops\\
GF11 &	1992 & 	Weingarten	&    PC/AT &     Weitek &	11 GFlops \\
\noalign{\smallskip}\hline
\multicolumn{2}{l}{*not for lattice QCD}\\
\end{tabular}
\end{center}
\end{table}

\begin{figure}
  \begin{center}
   \includegraphics[bb= 0 0 842 595,width=\textwidth]{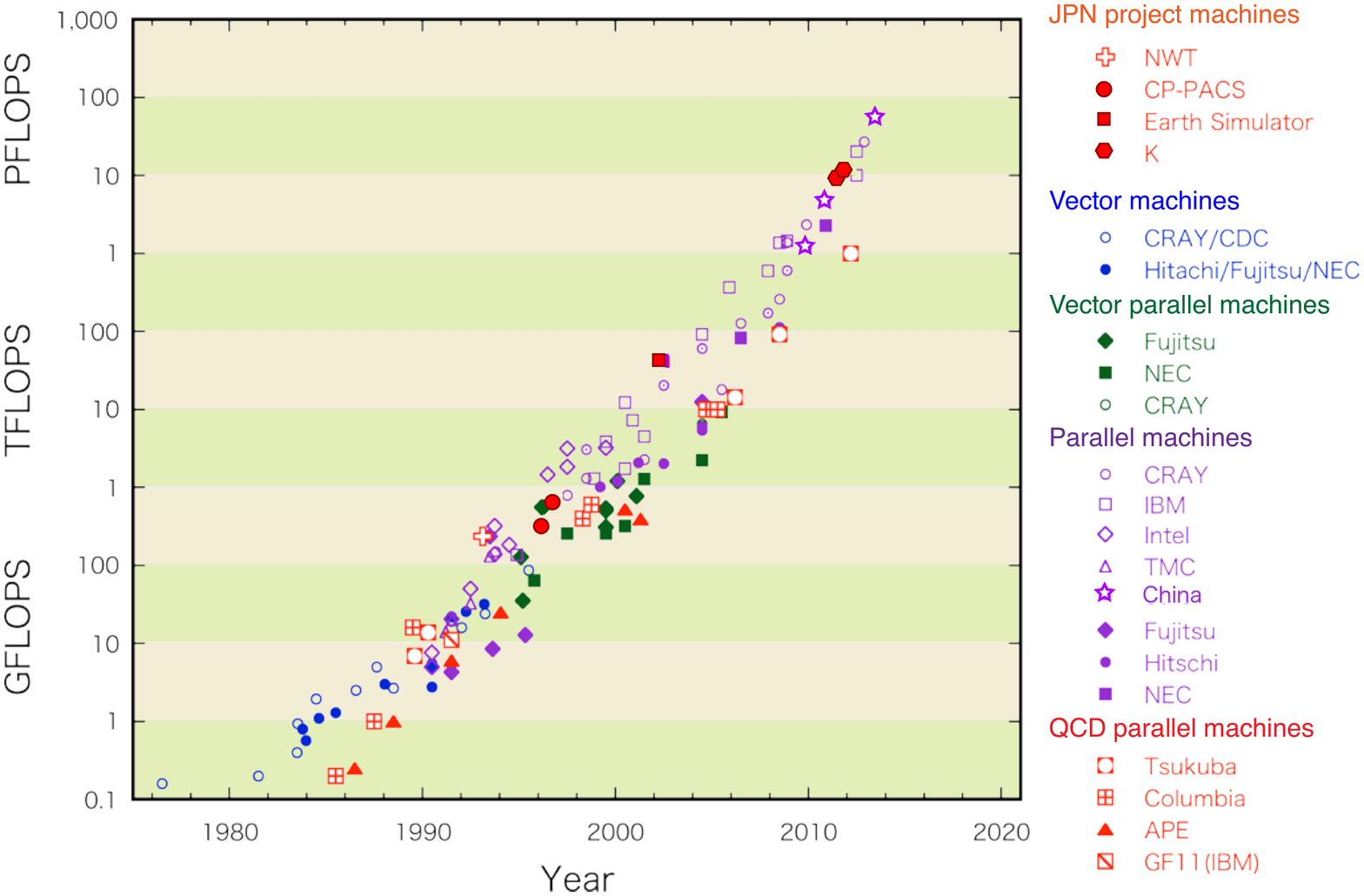}
  \end{center}
  \caption{Peak speed of supercomputers.  The circle at the bottom left is CRAY-1.  Red fancy symbols show parallel machines developed for QCD.    Color online. }
  \label{fig:computerspeed}
\end{figure}

The famous CRAY-1 vector supercomputer came on the market in 1976.  Vector supercomputers developed rapidly, and dominated the market in the 70' and 80's.  However, the progress of parallel computers was even faster.  By the end of the 80's, parallel computers  caught up and even overtook vector computers in speed.  We can observe this trend by tracking blue and green symbols (vector computers) and red and violet symbols (parallel computers)  from the 80's to early 90's in Fig.~\ref{fig:computerspeed}. 

\begin{table}
\caption{Parallel computers developed for lattice QCD in the 90's and later. ''year" marks the completion date.}
\label{tab:90s}  
\begin{center}
\begin{tabular}{llllll}
\hline\noalign{\smallskip}
name & year & authors  & CPU & vendor & peak speed  \\
\noalign{\smallskip}\hline\noalign{\smallskip}
APE100 &	1994&	APE Collab.  &   custom	&	 --	&	0.1Tflops\\
CP-PACS&	1996&	Iwasaki et al&	     PA-RISC	&	Hitachi(SR2201)	&	0.6TFlops\\
QCDSP&		1998&	Christ et al	&     TI DSP	& 	 -- 	&	0.6TFlops \\
APEmille &	2000&	APE Collab.  &   custom	&	 --	&	0.8Tflops\\
QCDOC &	2005&	Christ et al	&     PPC-based &	IBM(BG/L) &10TFlops\\
PACS-CS&2006&Ukawa et al&	  Intel  Xeon	& 	Hitachi	&	14TFlops\\ 
QCDCQ&2011&	Christ et al	&     PPC-based	& 	IBM(BG/Q)	&500TFlops\\
QPACE &		2012&	Wettig et al	&     PowerXCell &	  --	&	200TFlops\\
\noalign{\smallskip}\hline
\end{tabular}
\end{center}
\end{table}

In Table~\ref{tab:90s} we list the parallel computers developed in the 90's and later for lattice QCD.  Big success continued with CP-PACS in Japan, which occupied the top position in the Top 500 list of supercomputers in November 1996, and QCDSP in USA.  We observe an increasing involvement of major vendors such as Hitachi and IBM.  This was necessary to secure advanced technology and computer building knowhow to develop those fast computers.  Probably the most well-known of this trend is the QCDOC project, which gave rise to the IBM BlueGene/L, and the QCDCQ project which ran parallel with the IBM BlueGene/Q development.  In this way lattice QCD has seriously influenced the development of parallel supercomputers for scientific computing.   Another trend one can observe in Table~\ref{tab:90s} is the use of commodity processors such as Intel Xeon for a quick machine building.  This was the option adopted by PACS-CS and QPACE.  

Today, the fastest supercomputers have reached the peak speed of $O(50)$ Pflops~\cite{top500}.  This has been achieved in two ways.  The K computer in Japan and Sequoia (BlueGene/Q) in USA connected $O(10^5)$ multi-core processors running at $O(100)$~Gflops.  Tianhe-2 in China and Titan in USA boosted the computing power by adding multiples of GPGPU's running at $O(1)$~Tflops to each node.  Further increase of computing speed faces a serious issue that the memory cannot supply data fast enough to the processing units and the power consumption is becoming too large for large systems. 
Serious effort is already going on, however, to overcome these problems. 

\subsection{Fermion problem and hybrid Monte Carlo method}
\label{sec:2-4}

Monte Carlo calculation for the gluon fields, though somewhat complicated by the SU(3) nature of the field variable,  is straightforward.  Calculations for the quark fields, on the other hand, cannot be directly put on a computer since quark fields are represented by anticommuting Grassmann numbers.  Instead one uses the identity~\footnote{Strictly speaking, this identity requires positivity of the Hermitian part of matrix $D$.  We shall not go into this technical detail and mention only that this can be guaranteed.}
\begin{eqnarray}
&&\int\prod_n d\overline{q}_n dq_n\exp\left(\sum_{n,m}\overline{q}_nD_{n,m}(U)q_m\right) \\
&=&\int\prod_n d\phi^\dagger_n d\phi_n\exp\left(-\sum_{n,m}\phi^\dagger_nD^{-1}_{n,m}(U)\phi_m\right),  
\end{eqnarray}
where $D$ represents the lattice Dirac operator and $\phi_n$ represents a bosonic field with 4 Dirac  and 3 color indices to rewrite 
\begin{eqnarray}
\label{eq:fullQCD}
Z_{\rm QCD}=\int\prod_{n\mu}dU_{n\mu}\int\prod_n d\phi^\dagger_n d\phi_n \exp\left(S_{\rm gluon}-\sum_{n,m}\phi^\dagger_nD^{-1}_{n,m}(U)\phi_m\right).
\end{eqnarray}
with which the fundamental variables $U_{n\mu}$'s and $\phi_n$'s are all bosonic.

The inverse $D^{-1}_{n,m}(U)$ is a non-local quantity.  A change of a $\phi_n$ spreads across the lattice through the inverse.  Therefore  preparing a trial configuration whose acceptance can be controlled is not straightforward.  A number of methods were developed in the 80's including the  micro canonical~\cite{CallawayRahman1982,PolonyiWyld1983} and Langevin~\cite{UkawaFukugita1985} methods.  The latter was also explored by the group at Cornell including Wilson~\cite{Wilson1985}.  The standard method has settled on the hybrid Monte Carlo (HMC) method proposed in 1987~\cite{DuaneKennedyPendleton1987}, which we now discuss.  

The first step of HMC is to introduce an SU(3)-algebra valued momentum $P_{n\mu}$ conjugate to $U_{n\mu}$, and rewrite the path integral (\ref{eq:fullQCD}) of full QCD as a partition function of a fictitious classical system of $U_{n\mu}$'s and $P_{n\mu}$'s as 
\begin{eqnarray}
Z_{\rm QCD}=\int\prod_{n\mu}dP_{n\mu}\prod_{n\mu}dU_{n\mu}\int\prod_n d\phi^\dagger_n d\phi_n \exp\left(-H\right),
\end{eqnarray}
where $H$ is a fictitious Hamiltonian defined by 
\begin{eqnarray}
H=\frac{1}{2}\sum_{n\mu}\mathrm{Tr}\left(P_{n\mu}^2\right)-S_{\rm gluon}(U)+ \sum_{n,m}\phi^\dagger_nD^{-1}_{n,m}(U)\phi_m.
\end{eqnarray}

We now wish to generate a set of configurations distributed according to the weight $\exp(-H)/Z_{\rm QCD}$ by a Monte Carlo procedure.  For this purpose, one introduces a fictitious time $\tau$ conjugate to the Hamiltonian $H$. 
Starting with a given configuration of $U_{n\mu}$ and $\phi_n$ at $\tau=0$, we solve Hamilton's equations for  the canonical pair, 
\begin{eqnarray}
\frac{d}{d\tau}U_{n_\mu}&=&iU_{n\mu}P_{n\mu},\\
\label{eq:hamilton}
\frac{d}{d\tau}P_{n\mu}&=&\frac{\partial}{\partial U_{n\mu}}S_{gluon}(U) \nonumber \\ 
&+&\sum_{n,k,l,m}\phi_n^\dagger D_{n,k}^{-1}(U)\frac{\partial}{\partial U_{n\mu}}D_{k,l}(U)D_{l,m}^{-1}(U)\phi_m
\end{eqnarray}
over some interval of  $\tau$.  The configuration at a final time $\tau$ is used as a new configuration in the Monte Carlo procedure.  

In numerical implementations, a continuous fictitious time evolution is discretized with a finite step size $\delta\tau$.  Since the Hamiltonian is no longer conserved, the configuration generated after a number of steps $\tau_n=n\delta\tau$ suffers from a bias.  This is corrected by accepting or rejecting the generated configuration according to the Metropolis probability $P_{\rm acc}={\rm min}(1, \exp(-\delta H))$ where $\delta H=H(P(\tau_n), U(\tau_n))-H(P(0), U(0))$ is the difference of Hamiltonian over the trajectory of length $\tau_n=n\delta\tau$. 

Hybrid Monte Carlo is an elegant method.  It is (i) exact,  (ii) allows control of acceptance, {\it i.e.,} the probability that a trial configuration is accepted, through the magnitude of $\delta\tau$, and (iii) solving  Hamltion's equations can be executed in a parallel fashion.  However, at every discretized step $\delta\tau$ in updating the fields according to the Hamilton's equation, it requires the inverse of the lattice Dirac operator of form $x_n=\sum_m D_{n,m}^{-1}(U)\phi_m$ for given $U_{n\mu}$ and $\phi_n$.  

The inverse can be obtained by solving the lattice Dirac equation, 
\begin{eqnarray}
\label{eq:Dirac}
\sum_m D_{n,m}(U)x_m=\phi_n. 
\end{eqnarray}
This is a linear equation for a large but sparse matrix $D_{n,m}$, which can be obtained by iterative algorithms such as the conjugate gradient method.  The number of iterations $N_{\rm iter}$ needed to reach an approximate solution is controlled by the condition number $\kappa(D)$ of the matrix $D$.  It is given by $\kappa(D)=\lambda_{\rm max}/\lambda_{\rm min}$ where $\lambda_{\rm max}$ and $\lambda_{\rm min}$ are the maximum and minimum value of the eigenvalues of $D$.   Typically, the iteration number $N_{\rm iter}$ is inversely proportional to the condition number.  Since the minimum eigenvalue for the lattice Dirac operator is of the order of quark mass $m_qa$ in lattice units, and the maximum eigenvalue  is $O(1)$, one finds $N_{\rm iter}\propto 1/m_qa$.  

Among the six types of quarks known experimentally, the lightest up and down quarks have masses of the order of a few MeV.  This is three orders of magnitude smaller than the typical hadronic scale of 1 GeV.  The condition number of the lattice Dirac operator for these quarks is large, and therefore, hybrid Monte Carlo simulation slows down considerably as one tries to approach the physical values of quark masses.  

\subsection{Physical point calculation}
\label{sec:2-5}

\begin{figure}
  \begin{center}
   \includegraphics[bb= 0 0 620 431,width=0.5\textwidth]{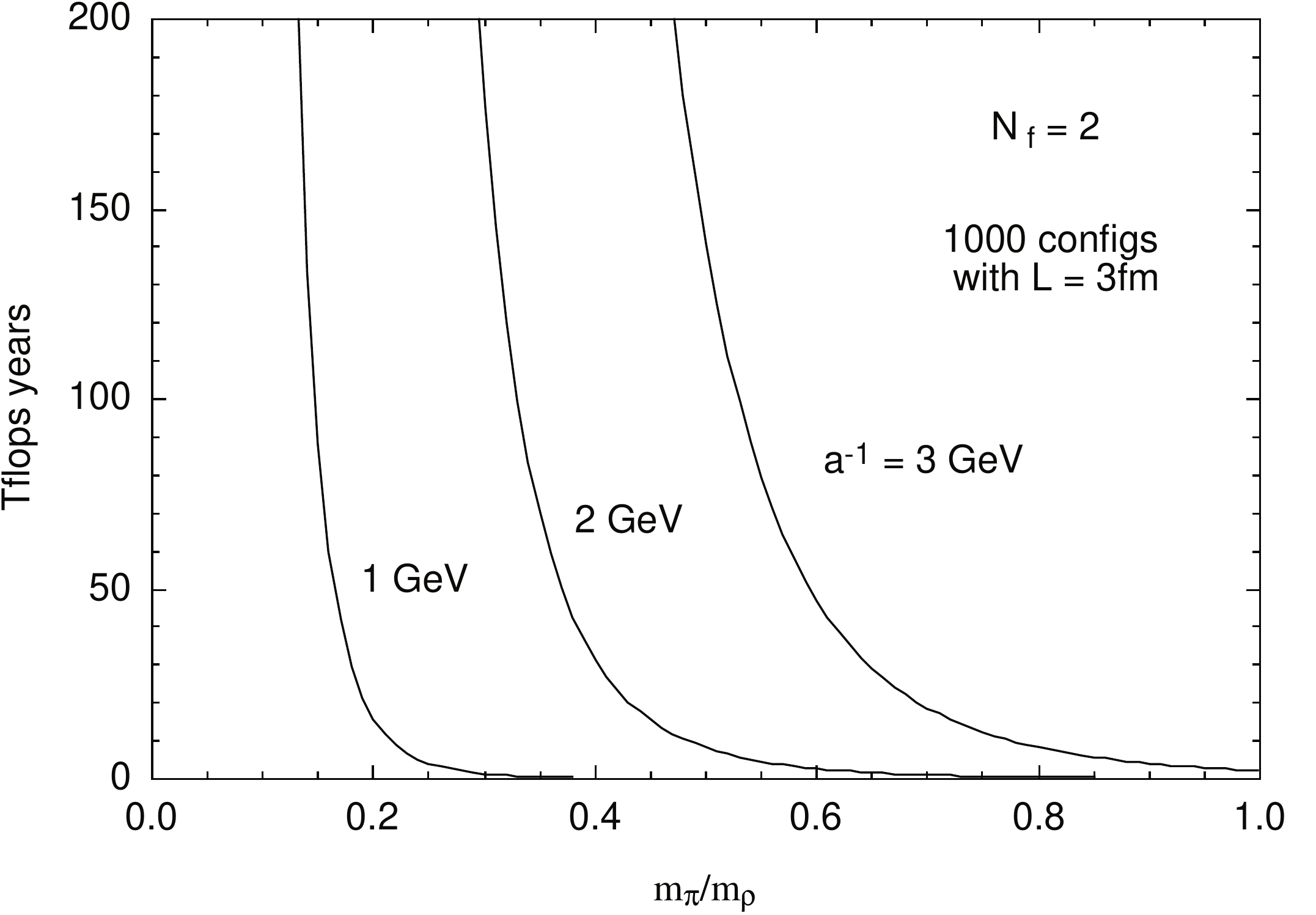}
  \end{center}
  \caption{Computational cost for generating 1000 independent configurations for $N_f=2$ flavor full QCD as a function of $m_\pi/m_\rho$ whose physical value equals $0.18$.  The three curves correspond to the lattice spacing $a^{-1}=1,2,3$~GeV.  Units is Tflops$\times$years, {\it i.e.,} 1 year of time with the average speed of 1 Tflops. From Ref.~\protect{\cite{Ukawa2001}}. }
  \label{fig:Berlinwall}
\end{figure}

Lattice QCD simulations including quark effects through hybrid Monte Carlo algorithm developed rapidly toward the end of the 90's.  By the turn of the century, there was enough experience accumulated to empirically estimate how much computing is needed to calculate observables with a quotable error for a given quark mass.  In Fig.~\ref{fig:Berlinwall} is shown a typical plot~\cite{Ukawa2001}, taking the case of $N_f=2$ flavor simulations for a lattice box of a size 3~fm, a minimum value to contain a hadron such as a nucleon within the box.  The vertical axis shows the amount of computing in units of Tflops$\times$year, {\it i. e.},  1 year of running a computer executing 1~Tflop$=10^{12}$ floating point operations per second.  The horizontal axis is the ratio of $\pi$ meson to $\rho$ meson masses, which varies with up and down quark masses and equals $m_\pi/m_\rho=0.18$ in experiment.  The three curves correspond to the inverse lattice spacing $1/a=1, 2, 3$~GeV, with $1/a=2$~GeV or larger being needed for a reliable continuum extrapolation.  We observe a sharp rise of the curves toward the physical value of $m_\pi/m_\rho=0.18$.  This is a critical slowing down of the hybrid Monte Carlo algorithm, primarily arising from the slowdown of the Dirac solver toward $m_qa\to 0$. 

The rapid increase presented a major problem for lattice QCD simulations.  Without overcoming this problem, one could only compute at quark masses much heavier than the physical values.  The results had to be extrapolated to the physical point, but such an extrapolation involved large systematic errors because of the existence of logarithmic terms of the form $\sim m_q \log m_q/\Lambda$ expected at vanishing quark mass. 

The difficulty was resolved in the middle of the 2000's~\cite{Luescher2005}.  Since the force in the Hamilton's equation (\ref{eq:hamilton}) involves $D^{-1}_{nm}(U)$, there are contributions coming from   the short-distance neighborhood of the link $n\mu$ being updated and from those further away.  It is possible to rewrite the quark determinant $\det D(U)$ such that these two types of contributions are separated.  It was shown in \cite{Luescher2005} that,  decomposing the lattice $\Lambda$ into a set of sub-lattices $\Lambda_i, i=1,2,\cdots$,  one can write 
\begin{eqnarray}
\det D(U) = \prod_i \det D_i(U)\cdot \det R(U),
\end{eqnarray}
where $D_i(U)$ is the Dirac operator restricted to a sub-lattice $\Lambda_i$ and $R$ couples sites belonging to different sub-lattices.  Rewriting each determinant factor on the right hand side as a bosonic integral, one  can write the force term as 
\begin{eqnarray}
F_{n\mu}=F_{n\mu}^{\rm gluon}+F_{n\mu}^{\rm UV}+F_{n\mu}^{\rm IR},
\end{eqnarray}
where $F_{n\mu}^{\rm gluon}$ is the term coming from the gluon action, $F_{n\mu}^{\rm UV}$ those arising from $D_i(U)$'s, and hence represents short-distance contributions,  and  $F_{n\mu}^{\rm IR}$ is the term from $R(U)$ giving long-distance contributions.  
 
\begin{figure}
  \begin{center}
   \includegraphics[bb=0 0 525 383,width=0.5\textwidth]{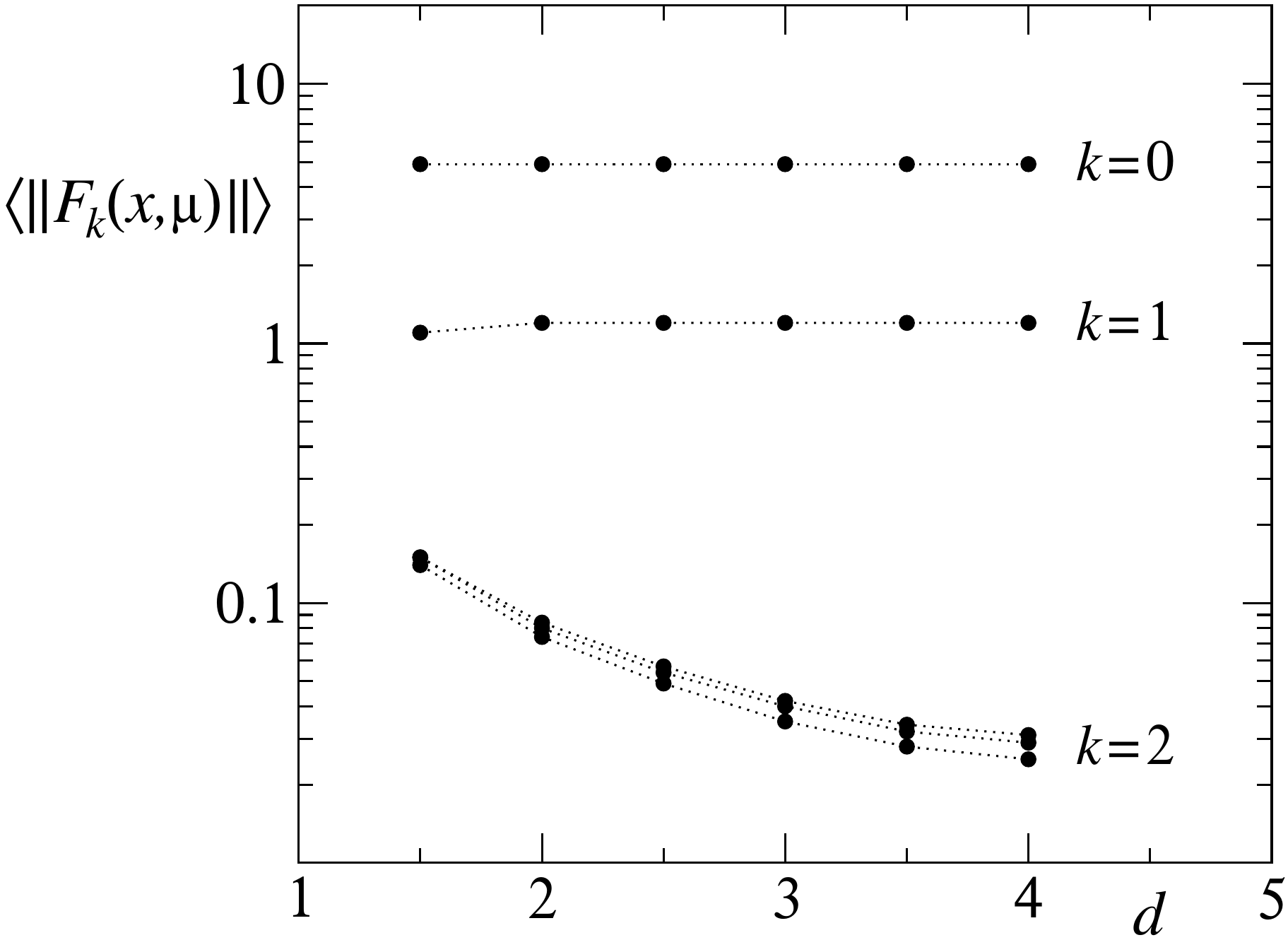}
  \end{center}
  \caption{Magnitude of the force term $F_{n\mu}^{\rm gluon} (k=0)$, $F_{n\mu}^{\rm UV}(k=1)$, and $F_{n\mu}^{\rm IR}(k=2)$ as a function of distance from sub lattice boundary.  From Ref.~\protect{\cite{Luescher2005}}. }
  \label{fig:luescher}
\end{figure}

 In Fig.~\ref{fig:luescher} we show the magnitude of the three terms observed in a hybrid Monte Carlo run~\cite{Luescher2005}.  There is  a clear separation of magnitude for the gluon, and UV and IR quark contributions.  Therefore, the step size $\delta\tau_{\rm IR}$ for the IR part $F_{n\mu}^{\rm IR}$ can be taken much larger than  $\delta\tau_{\rm UV}$ for the UV part, which in turn can be taken much larger than $\delta\tau_{\rm gluon}$ for the gluon part.  An evolution with different step sizes for different parts of the force can be realized by a multi-time step integrator originally developed in \cite{SextonWeingarten}.

Since inverting the long-distance part of the operator $R$ is the computationally most intensive, using different step sizes for the three parts of the force lead to  acceleration of the evolution by a large factor of order $\delta\tau_{\rm IR}/\delta\tau_{\rm gluon}\approx 10-50$.  Simulations incorporating such an acceleration technique were first carried out in the late 2000's~\cite{PACS-CS2009}, and reached the physical quark masses for the up and down quarks.   

The separation of UV and IR parts of the quark force can be realized in a different manner~\cite{Hasenbusch}  by rewriting the Dirac operator for a quark mass $m_0$ as a product of ratios of successively heavier masses, 
\begin{eqnarray}
D(U, m_0)=\prod_{i=0}^{N}\left(D(U,m_i)D^{-1}(U, m_{i+1})\right)\cdot D(U, m_N),
\end{eqnarray}
where $m_{i+1}$ is taken larger than $m_i$.  Qualitatively speaking, one is separating the contributions to the force according to the eigenvalues in the range $[m_i, m_{i+1}]$ with $m_{N+1}$ equal to the largest eigenvalue.  

A somewhat different method to accelerate the hybrid Monte Carlo algorithm is provided by the rational hybrid Monte Carlo algorithm~\cite{Horvathetal1999}.  One writes 
\begin{eqnarray}
\det D =\left[\det D^{1/n}\right]^n=\int\prod_{i=1}^n d\phi^{i\dagger}d\phi^{i}\exp\left[-\sum_{i=1}^n\phi^{i\dagger}D^{-1/n}\phi^{i}\right]
\end{eqnarray} 
for some positive integer $n$, and applies a rational approximation to the fractional power $x^{-1/n}$.  
If $\kappa$ is the condition number of $D$, the n-th root $D^{1/n}$ will have a smaller condition number $\kappa^{1/n}$.  The magnitude of the force, compared with that of the hybrid Monte carlo method, will be reduced by a factor $\kappa^{1-1/n}$.  Hence the step size can be increased by the corresponding factor, leading to an acceleration of the molecular dynamics evolution. 
 
A final comment concerns the iterative inversion of the lattice Dirac operator (\ref{eq:Dirac}).  Since the increase of the number of iterations toward small quark masses is the cause of the slowdown of the hybrid Monte Carlo algorithm, an ultimate optimization of the algorithm would be realized if one could remove the critical slowdown.  This has recently been achieved by understanding the modes corresponding to the small eigenvalues of the lattice Dirac operator.   One can either construct these modes explicitly and "deflate" ({\it i.e.}, remove) them from the inversion~\cite{Luescher2007},  or  employ multi-grid techniques to adaptively generate and incorporate these modes in the inversion~\cite{Babichietal2010,Frommeretal2013}.  

 Over the years, the improvements as described here plus the development of an increasingly powerful computers have made it possible to carry out lattice QCD calculations at the physical quark masses.  Such calculations are now routinely done.  This is an impressive achievement.  It also carries an aesthetic appeal; with the physical quark masses, one is no longer simulating, but rather calculating,  the physical processes of the strong interactions as they are taking place in our universe. 

\section{Physics results}
\label{sec:3}
\subsection{Light Hadron spectrum}
\label{sec:3-1}
\subsubsection{Mass spectrum of light mesons and baryons}
\label{sec:3-1-1}

Since the masses of hadrons are dynamical quantities, whether lattice QCD can quantitatively  explain the experimentally known mass spectrum provides a stringent test of the validity of QCD at low energies.  Furthermore, the success of such a calculation forms the basis on which the reliability of lattice QCD predictions of other physical quantities are to be built.  For these reasons, the calculation of the hadron mass spectrum, in particular those of the ground states of light mesons and baryons composed of up, down, and strange quarks, has been pursued since the beginning of lattice QCD and numerical simulations. 

A precise  determination of the mass spectrum has to control a number of sources of errors.  They are (i) statistical error due to the Monte Carlo nature of the calculation, (ii) systematic error due to extrapolation to the physical quark masses, (iii) systematic error due to finite lattice sizes, (iv) systematic error due to finite lattice spacings.  In addition, until the late 90's, most calculations were carried out with the so-called quenched approximation in which effects of quark-antiquark pair creation and annihilation are entirely ignored by dropping the quark determinant $\det D$ from the path integral.  This was due to a high cost of including the quark effects into the simulations.  

\begin{figure}
  \begin{center}
      \includegraphics[bb=0 0 401 376,width=0.40\textwidth]{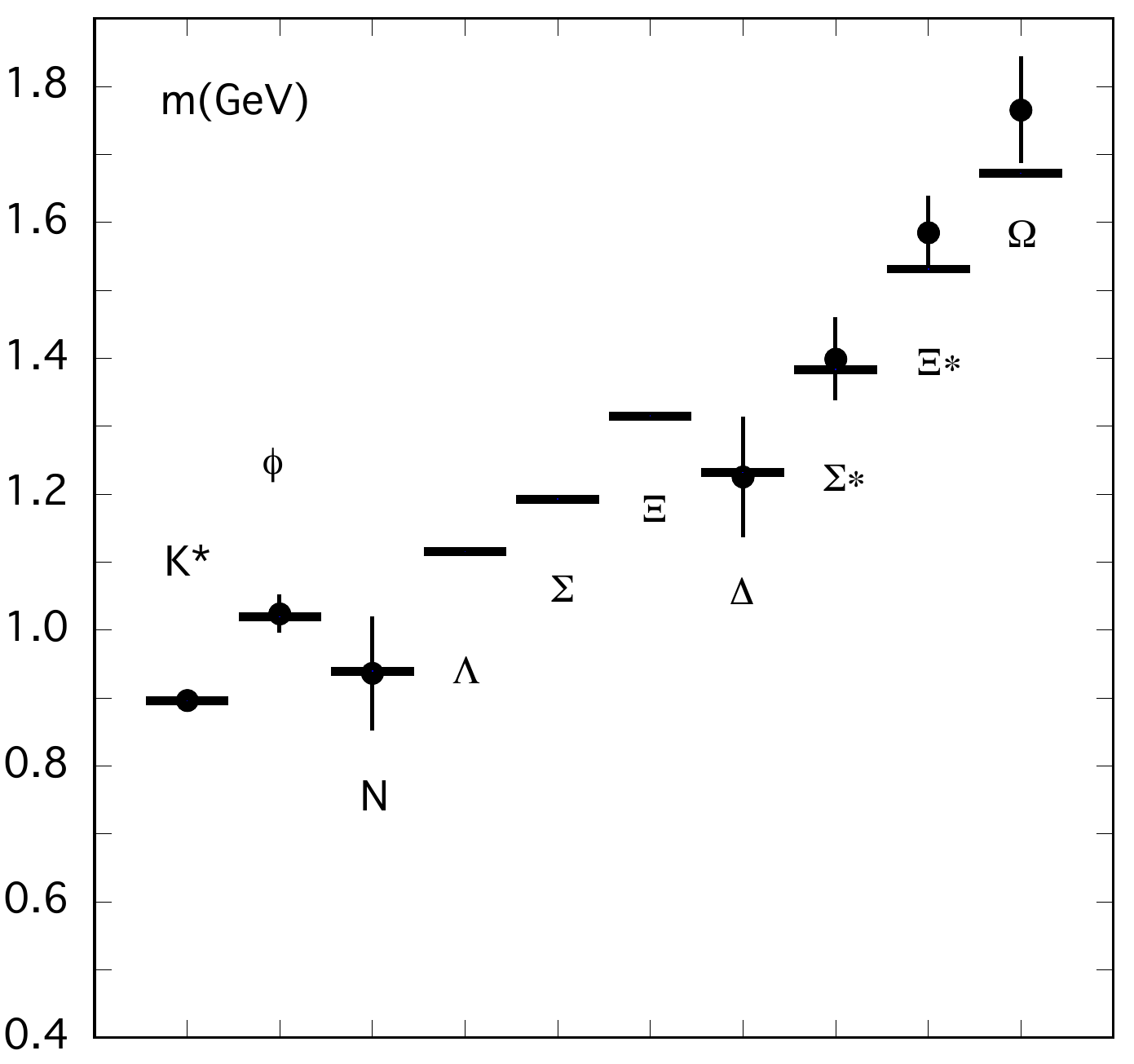}
      \hspace*{10mm}
      \includegraphics[bb=0 0 493 406,width=0.45\textwidth]{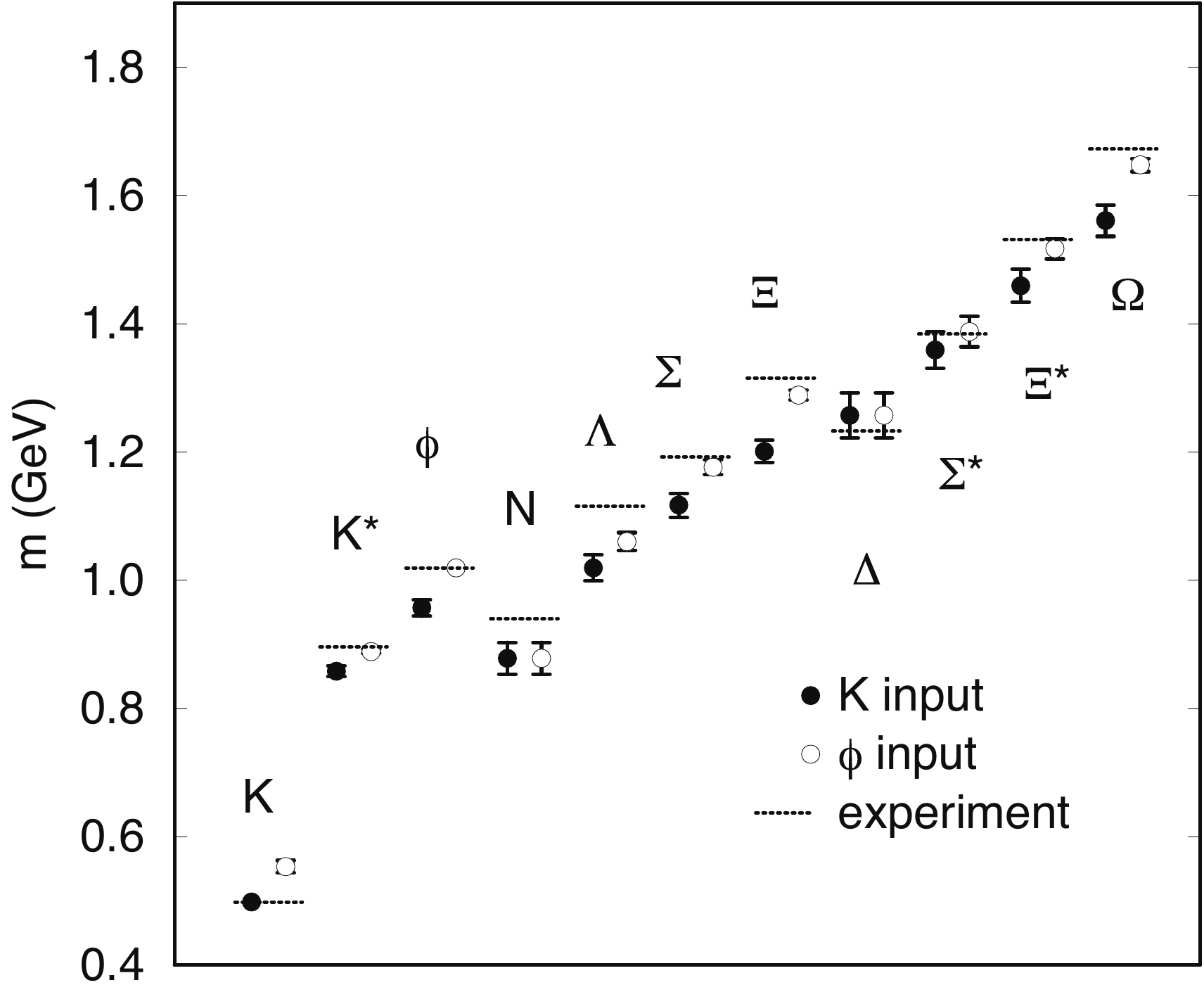}
     \end{center}
  \caption{The left panel shows the result of the first systematic quenched light hadron mass spectrum calculation~\protect{\cite{gf11}} published in 1993.  The right panel shows the definitive quenched result~\protect{\cite{CP-PACS}} reported in 2000. }
  \label{fig:quenched}
\end{figure}

The first systematic attempt at a precision calculation of the  light hadron spectrum was carried out by Weingarten and his colleagues within the quenched approximation in 1993~\cite{gf11}.  The calculation took 1 year of dedicated computer time on GF11 computer developed by him at IBM.  This was a landmark calculation in that a controlled extrapolation to the physical quark masses, to infinite volume and to zero lattice spacing, was systematically attempted.  
The masses of $\pi$, $\rho$ and $K$ mesons were employed to fix degenerate up and down, and strange quark masses.  The masses of $K^*$ and $\phi$ for the meson octet, nucleon $N$ for the baryon octet,  and $\Delta, \Sigma^*, \Xi^*$ and $\Omega$ for the baryon decuplet were obtained as predictions.  They are plotted by filled circles in the left panel of Fig.~\ref{fig:quenched}.  The masses of $\Lambda, \Sigma$ and $\Xi$ of the baryon octet were not reported. The horizontal bars show the experimental values.   We observe that the calculated values are consistent with experiment within one standard deviation.  However,  for baryons, a sizable magnitude of the errors of O(10$\%$) make it difficult to conclude if there is a precise agreement. 

A definitive calculation of the hadron spectrum within the quenched calculation followed in 2000~\cite{CP-PACS}.  This work  took half a year of  CP-PACS computer which was  55 times faster than GF11.  The results for the light meson and baryon ground states are shown in the right panel of Fig.~\ref{fig:quenched}.  As one clearly sees there,  the quenched spectrum systematically deviates from the experimental spectrum. If one uses the K meson mass $m_K$ as input to fix the strange quark mass (filled data points), the vector meson masses $m_{K^*}$ and $m_\phi$ are smaller by 4\%($4\sigma$) and 6\% ($5\sigma$), the octet baryon masses are smaller by 6\% -- 9\% ($4-7\sigma$), and the decuplet mass splittings are smaller by 30\% on average.  Alternatively, if the $\phi$ meson mass $m_\phi$ is employed (open circles), $m_{K^*}$ agrees with experiment within 0.8\% ($2\sigma$) and the discrepancies for baryon masses are much reduced. However, $m_K$ is larger by 11\% ($6\sigma$). In other words, there is no way to match the entire spectrum beyond 5 to 10\% precision in quenched QCD. 

The CP-PACS calculation heralded the end of the era of quenched calculations. Efforts toward full QCD simulations, which had already been taking shape in the late 90's, intensified.   A first systematic calculation in full QCD was made by the CP-PACS Collaboration with dynamical up and down quarks ($N_f=2$)~\cite{CP-PACS2000}. With the algorithmic development in the middle of the 2000's, which we described in Sec.~\ref{sec:2-5}, calculations around the physical quark masses became possible.  The PACS-CS Collaboration carried out $N_f=2+1$ flavor calculations in which the strange quark mass was taken close to the physical value and the pion mass was decreased down to $m_\pi=155$~MeV as compared to the physical value of 135~MeV~\cite{PACS-CS2009}.  

\begin{figure}[t]
  \begin{center}
   \includegraphics[bb=0 0 703 454,width=0.6\textwidth]{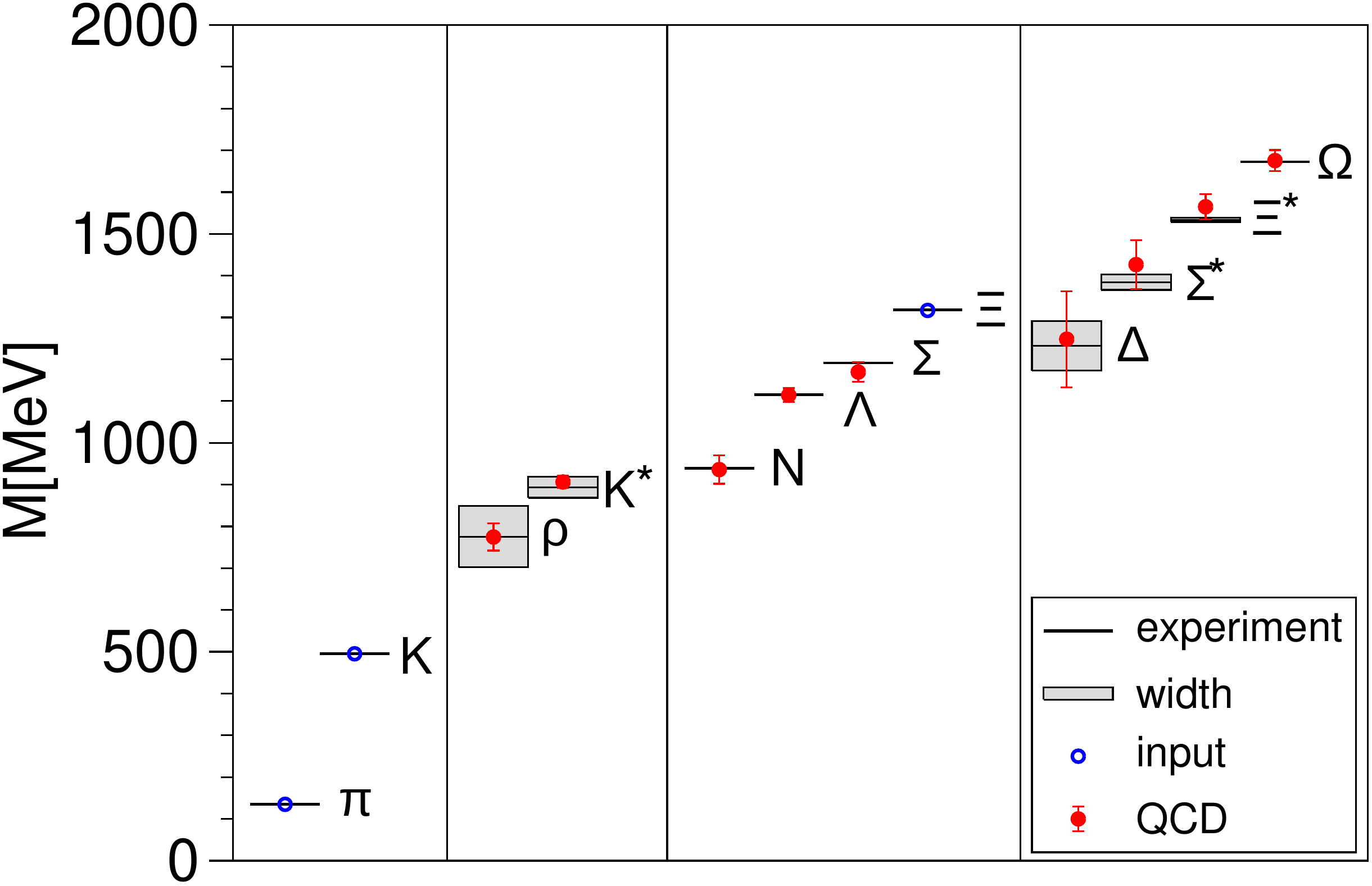}
  \end{center}
  \caption{Light hadron mass spectrum in $N_f=2+1$ flavor QCD~\cite{BMW2008} reported in 2008.  Masses of $\pi$, $K$ and $\Xi$ are used to fix the degenerate up-down and strange quark masses as well as scale. Boxes of shaded grey show resonance masses and widths.  Color online.}
  \label{fig:BMW}
\end{figure}

Finally, the BMW Collaboration carried out an infinite volume and continuum limit extrapolated calculation in 2008~\cite{BMW2008}.  We reproduce their results  in Fig.~\ref{fig:BMW}.  There is good agreement with the experimental spectrum within the errors of 2--5\% except for $\Delta$ for which the error is 13\%. 

Some of hadrons, including $\rho$ and $\Delta$, are resonances which decay for physical quark masses and infinite volume.  A general relation between energy levels of two body states at finite volume and scattering phase shift, from which resonance parameters can be extracted, was established by L\"uscher some time ago~\cite{Luescher1991}.  The BMW Collaboration assumed the Breit-Wigner form for the phase shift, and used this relation to correct the measured masses for the effects of decay coupling.   

A more rigorous approach in which the phase shift is first extracted from measured energy levels and L\"uscher's formula,  and resonance masses and widths are subsequently determined from the phase shift, was pioneered in \cite{Ishizuka2007}.  Because of severe computational cost, it will take more time before physical point calculations can treat resonances numerically precisely in this way.  

\subsubsection{Isospin splittings of light hadron masses}

The isospin multiplets of hadrons exhibit small mass differences of a few MeV.  These tiny effects are nonetheless very important to understand our universe.  For example, neutron with a mass $m_n=939.565379(21)$~MeV is heavier than proton of a mass $m_p=938.272046 (21)$~MeV by $m_n-m_p = 1.2933322(4)$~MeV.  Because of this difference, neutron undergoes a $\beta$ decay with a mean life of $880.3(1.1)$~sec, which has important consequences in nucleosynthesis after the Big Bang, and the composition of nuclei as we see them today.  

Since isospin mass splittings arise from the mass difference of up and down quarks,  $N_f=1+1+1$ simulations with different masses for up, down, and strange quarks become necessary.  Once we take the quark mass difference into account, one also likes to consider the electromagnetic effects, since  the magnitude of the effect, expected at the order of a few MeV, is similar to those arising from the up-down quark mass difference.  

A pioneering work on isospin breaking effects was carried out in the mid-90's~\cite{DuncanEichtenThacker1996}.  The RBC Collaboration rekindled interest by pointing out its importance~\cite{Blumetal2010}. Both quenched QED~\cite{Blumetal2010} and full QED~\cite{Ishikawaetal2012,PACS-CS2012} have been attempted. 
The most recent calculation has been reported by the BMW Collaboration~\cite{BMW2014}.  They carried out $N_f=1+1+1+1$ QCD and QED simulations with independent up, down, strange, and charm quark masses and three values of QED coupling  for a variety of lattice sizes and spacings.  A careful analysis of finite size effects due to the infinite range of photon was made.   The infinite volume and continuum limit extrapolation was carried out.  The final result is 
\begin{eqnarray}
m_n-m_p = +1.51(28) {\rm ~MeV}
\end{eqnarray}
in good agreement with experiment.  Treating the hadron mass differences to first order in $m_u-m_d$ and QED coupling $\alpha$, they could separate QCD and QED contributions with the result, 
\begin{eqnarray}
m_n-m_p = +2.52(30)_{\rm QCD} - 1.00(16)_{\rm QED} {\rm ~MeV}.
\end{eqnarray}
There is a delicate cancellation between the QCD and QED effects before the final number settles on the experimental value. 
 
\subsection{Fundamental constants of the strong interaction}
\label{sec:3-2}
\subsubsection{Quark masses}
\label{sec:3-2-1}

\begin{table}[b]
\caption{Light quark masses in MeV units in the $\overline{\rm MS}$ scheme at $\mu=2$~GeV.}
\label{tab:quarkmasses}  
\begin{center}
\begin{tabular}{lllllll}
\hline\noalign{\smallskip}
year 	&   &action & $\frac{m_u+m_d}{2}$  & $m_u$ 	& $m_d$ 	& $m_s$ \\
\hline\noalign{\smallskip}
\multicolumn{4}{l}{quenched QCD}&\multicolumn{3}{r}{${}^{1)}$ $m_K$ input ${}^{2)} $ $m_\phi$ input }\\
\hline\noalign{\smallskip}
2000 	& CP-PACS~\cite{CP-PACS}	& Wilson 	& 4.57(18) 	&	--	&	--	& 115.6(2.3)$^{1)}$ \\
		&						&		&			&		&		& 143.7(5.8)$^{2)}$\\
		\noalign{\smallskip}\hline\noalign{\smallskip}
\multicolumn{4}{l}{$N_f=2$ QCD}&\multicolumn{3}{r}{${}^{1)}$ $m_K$ input ${}^{2)} $ $m_\phi$ input }\\
\hline\noalign{\smallskip}
2000 	& CP-PACS~\cite{CP-PACS2000}	& Wilson 	& $3.45^{+0.14}_{-0.20}$ 	&	--	&	--	& $89^{+3}_{-6}$$\,\,^{1)}$ \\
		&						&		&			&		&		& $90^{+5}_{-10}$$^{2)}$\\
		\noalign{\smallskip}\hline\noalign{\smallskip}
\multicolumn{7}{l}{$N_f=2+1$ QCD}\\
\hline\noalign{\smallskip}
2010 	& MILC~\cite{MILC09}		& staggered & 3.19(18)	& 1.96(14) & 4.53(32) & 89.0(4.8) \\
2010		& BMW~\cite{BMW2010}		& Wilson	    & 3.469(67)	& 2.15(11) & 4.79(14) & 95.5(1.9) \\
2012		&RBC/	&  domain & 3.37(12) & --	& --	& 92.3(2.3) \\	
		&UKQCD~\cite{RBCUKQCDDomainwall2013} & wall\\
\hline\noalign{\smallskip}
\multicolumn{7}{l}{Reviews of Particle Physics}\\
\hline\noalign{\smallskip}
1998		& PDG~\cite{PDG1998}		&		    & 2--6	& 1.5--5	& 3--9  & 60--170 \\
2012		& PDG~\cite{PDG2012}		&		    & $3.5^{+0.9}_{-0.2}$ &$2.3^{+0.7}_{-0.5}$ & $4.8^{+0.5}_{-0.3}$ & $95\pm 5$ \\
\noalign{\smallskip}\hline
\end{tabular}
\end{center}
\end{table}

\begin{table}[t]
\caption{Heavy quark masses in GeV units in the $\overline{\rm MS}$ scheme at $\mu=m_h$.}
\label{tab:heavyquarkmasses}  
\begin{center}
\begin{tabular}{llllll}
\hline\noalign{\smallskip}
year 	&   &action & $m_c$ 	& $m_b$ 	& $m_t$ \\
\noalign{\smallskip}\hline\noalign{\smallskip}
\multicolumn{6}{l}{$N_f=2+1$ QCD}\\
\noalign{\smallskip}\hline\noalign{\smallskip}
2010 	& HPQCD~\cite{HPQCD2010}		& staggered & 1.273(6)	& 4.164(27) & --\\
\noalign{\smallskip}\hline\noalign{\smallskip}
1998		& PDG~\cite{PDG1998}		& 	&1.1--1.4		    & 4.1--4.4 & 173.8(5.2)\\
2012		& PDG~\cite{PDG2012}		& 	&1.275(25)	    & 4.18(3)  & 173.07(89) \\
\noalign{\smallskip}\hline
\end{tabular}
\end{center}
\end{table}

The determination of quark masses is a very important consequence of hadron mass spectrum calculations with lattice QCD.  Since quarks are confined unlike leptons, lattice QCD provides the only reliable way for finding the values of this set of fundamental constants of our universe.   

In Table~\ref{tab:quarkmasses} we list representative lattice results as well as estimates by Particle Data Group over the years.  Even as late as 1998, the Review of Particle Physics listed quite a wide band of values as shown in this table.  This exemplifies how uncertain quark masses were only a decade and a half ago.  Two years later, the CP-PACS quenched spectrum calculation narrowed the range considerably.  However, the limitation of quenched QCD is manifest in a large discrepancy of the strange quark mass depending on the input.  

The $N_f=2$ full QCD calculation~\cite{CP-PACS2000} including dynamical up and down quarks but treating strange quark in the quenched approximation showed that (i) quark masses are significantly smaller than indicated by the quenched results, and (ii) the discrepancy of strange quark mass depending on the input almost disappears.  

The recent results from $N_f=2+1$ full QCD  listed in Table~\ref{tab:quarkmasses} covers 3 types of fermion actions.  All three calculations carry out infinite volume and continuum extrapolations, albeit the degree of sophistication differs among the three.  The separate values of up and down quark masses are estimated using additional input on isospin breaking such as the $K^0-K^+$ mass difference and estimations on electromagnetic effects.  The three sets of results are reasonably consistent.  The 2012 Review of Particle Physics values reflect this progress.  

In Table~\ref{tab:heavyquarkmasses} we list the masses of charm and bottom quarks as determined by lattice QCD, together with those in Reviews of Particle Physics over the years.  We also list the value for the top quark for completeness. 

Lattice QCD is rapidly moving into the era when all three light quarks are treated independently, and dynamical effects of charm quark is also included.  It will be soon that such calculations yield a direct calculation of each of the quark masses with a few \% error. 

\subsubsection{Strong coupling constant}
\label{sec:3-2-2}

\begin{figure}[t]
  \begin{center}
   \includegraphics[bb=0 0 410 431,width=0.45\textwidth]{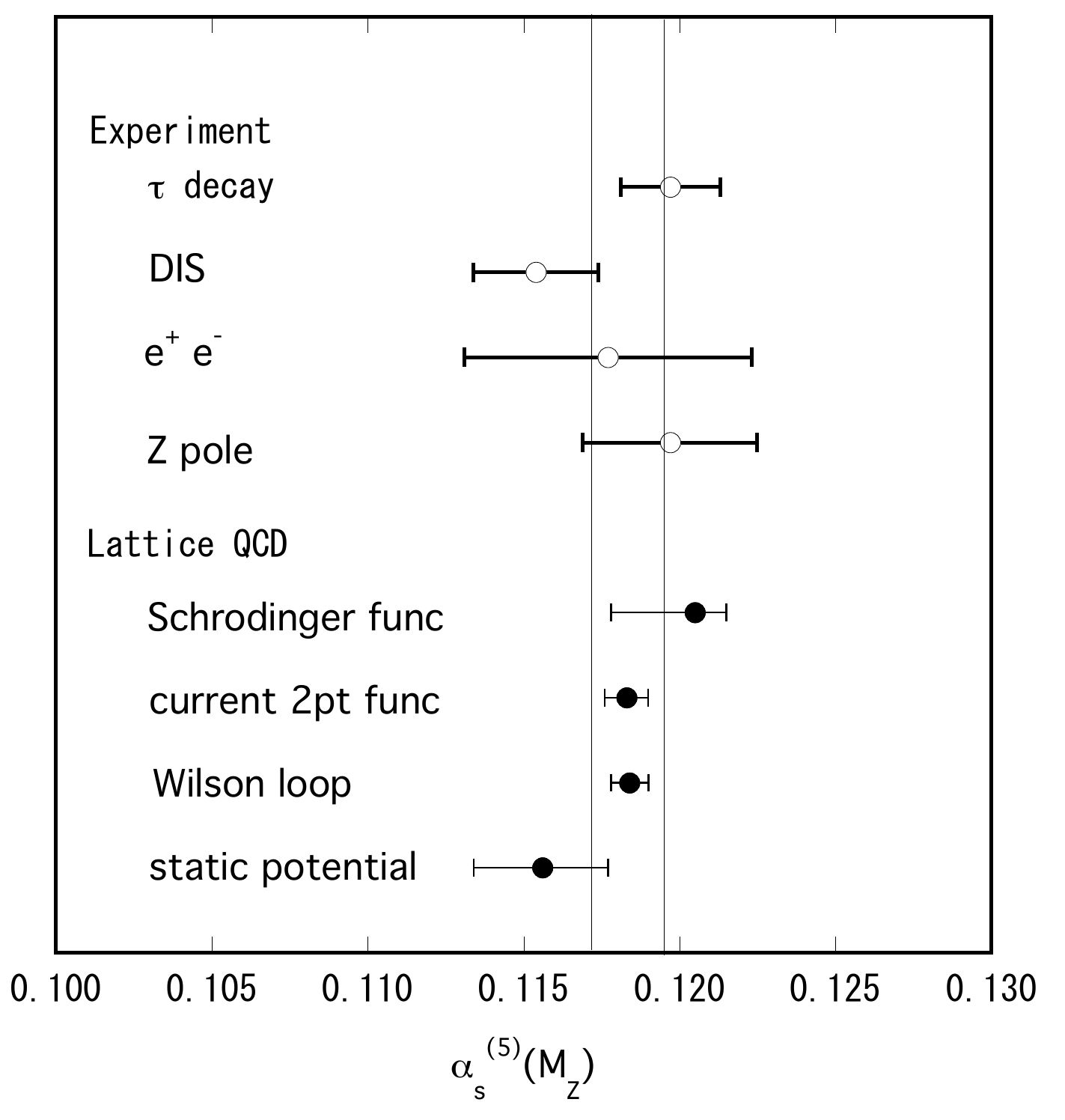}
  \end{center}
  \caption{Strong coupling constant $\alpha_s^{(5)}(M_Z)$ obtained from experiment and from lattice QCD.  Here experiment means determinations from scattering and decay data combined with perturbative QCD.  For further explanations, see text.}
  \label{fig:alphas}
\end{figure}
 
The value of the strong coupling constant $\alpha_s(\mu)=g^2(\mu)/4\pi$ defined in terms of the QCD coupling $g(\mu)$ at some prescribed scale $\mu$ is another fundamental constant of our universe, on a par with the fine structure constant $\alpha=e^2/4\pi$ for electromagnetism with the famous value $\alpha^{-1}=137.035 999 074(44)$.  

In Fig~\ref{fig:alphas} we plot the determinations obtained from scattering and decay data combined with perturbative QCD as of Fall 2013~\cite{PDG2013revision}, and compare them with lattice QCD values from recent $N_f=2+1$ calculations.  By convention, the scale is taken to be the Z boson mass $m_Z=91.1876(21)$~GeV, and five flavors of quarks excluding top is incorporated in the running of the coupling.  Lattice QCD determinations employ experimental quantities at low energies such as hadron masses and decay constants for fixing the scale.   The methods employed range from step scaling function using the Schr\"odinger functional by PACS-CS~\cite{PACS-CS2009}, current 2-point function and Wilson loops by HPQCD~\cite{HPQCD2010}, to static quark antiquark potential by Bazavov {\it et al}~\cite{Bazavov2012}.  The vertical band corresponds to the average over the four experimental determinations~\cite{PDG2013revision}, $\alpha_s^{(5)}(M_Z)=0.1183(12)$.  

The determinations with perturbative QCD still show some scatter and relatively large errors, and so do lattice QCD determinations depending on the method, the ones from current 2-point function and Wilson loops having the smallest error.  There is consistency with experiment at 1\% level, and  the precision of lattice determinations will steadily improve. 

\subsection{CP violation in the Standard Model}
\label{sec:3-3}
\subsubsection{CKM matrix elements}
\label{sec:3-3-1}

The connection between six types of quarks and CP violation is a salient feature of the Standard Model.  The complex phase characterizing CP violation is encoded in the Cabibbo-Kobayashi-Maskawa matrix $V$ which appears in the weak interaction Lagrangian as  
\begin{eqnarray}
L_{int}=ig\left( \begin{array}{ccc} \overline{u}, \overline{c}, \overline{t} \end{array} \right)_L\gamma_\mu
V \left( \begin{array}{c} d\\ s\\ b \end{array} \right)_LW_\mu^- + {\rm h.c.} 
\end{eqnarray}
In the Wolfenstein parametrization, the matrix takes the form,
\begin{eqnarray}
V=\left( \begin{array}{ccc} V_{ud}, V_{us}, V_{ub}\\
					V_{cd}, V_{cs}, V_{cb}\\
			             V_{td},  V_{ts}, V_{tb}\\
	\end{array}\right)
= \left( \begin{array}{ccc} 1-\lambda^2/2, \lambda, A\lambda^3(\rho-i \eta)\\
          			     -\lambda, 1-\lambda^2/2, A\lambda^2\\
			            A\lambda^3(1-\rho-i\eta), -A\lambda^2, 1\\
     			     	\end{array}\right)	+ O(\lambda^4),		     
\end{eqnarray}
with three real parameters $A, \lambda, \rho$ and an imaginary term  $i\eta$ characterizing CP violation.  

Constraining the CKM matrix requires matching experimental data on weak processes of hadrons to the theoretical expressions for the transition amplitudes which follow from the interaction Lagrangian above.  Since quarks interact strongly according to QCD,  the weak transition amplitudes are dressed by QCD corrections.  These corrections can be calculated by lattice QCD.  

One of weak processes which plays an important role is the CP violating mixing of neutral $K^0$ and $\overline{K}^0$ mesons.  The experimentally measured state mixing amplitude $\epsilon$ can be written as
\begin{eqnarray}
\epsilon=C\hat{B}_K\left\{\eta_1S(x_c){\rm Im}(\lambda_c^2)+\eta_2S(x_t){\rm Im}(\lambda_t^2)+2\eta_3S(x_c,x_t){\rm Im}(\lambda_c\lambda_t)\right\}.
\end{eqnarray}
Here $C$, $\eta_{1,2,3}$, $S(x, y)$, $S(x)=S(x,x)$ are known constants and functions, $x_q=m_q^2/m_W^2$ with $m_q$ and $m_W$ the mass of quark $q$ and W boson, and $\lambda_q=V_{qs}^*V_{qd}$ is the product of CKM matrix elements.  The K meson bag parameter $B_K$ is defined as 
\begin{eqnarray}
B_K=\frac{\left<\overline{K}^0\vert \left(\overline{s}\gamma_\mu d\right)_L\left(\overline{s}\gamma_\mu d\right)_L\vert K^0\right>}{\frac{8}{3}f_K^2m_K^2},
\end{eqnarray}
where the expectation value is taken for QCD, and $\hat{B}_K$ is its renormalization group invariant value.  We see then that the precision with which we can compute $B_K$ directly reflects in the precision in the determination of the CKM matrix elements through the products $\lambda_q=V_{qs}^*V_{qd}$.  

Similarly, for the mixing of $B_q^0$ and $\overline{B}_q^0$ mesons with $q=d$ or $s$, the oscillation frequency $\Delta m_q$ is proportional to the $B_q$ meson decay constant $f_{B_q}$ and the bag parameter $B_{B_q}$ defined by   
\begin{eqnarray}
B_{B_{q}}=\frac{\left<\overline{B}^0\vert \left(\overline{b}\gamma_\mu q\right)_L\left(\overline{b}\gamma_\mu q\right)_L\vert B^0\right>}{\frac{8}{3}f_{B_q}^2m_{B_q}^2}.
\end{eqnarray}
In addition, the decays of $B$ mesons in the leptonic ($B\to \ell\nu$) and semi-leptonic (such as $B\to\pi\ell\nu$ and $B\to D^*\ell\nu$) channels are used to constrain the matrix elements $\vert V_{ub}\vert$ and $\vert V_{cb}\vert$.

\begin{figure}[t]
  \begin{center}
   \includegraphics[bb=0 0 567 546,width=0.6\textwidth]{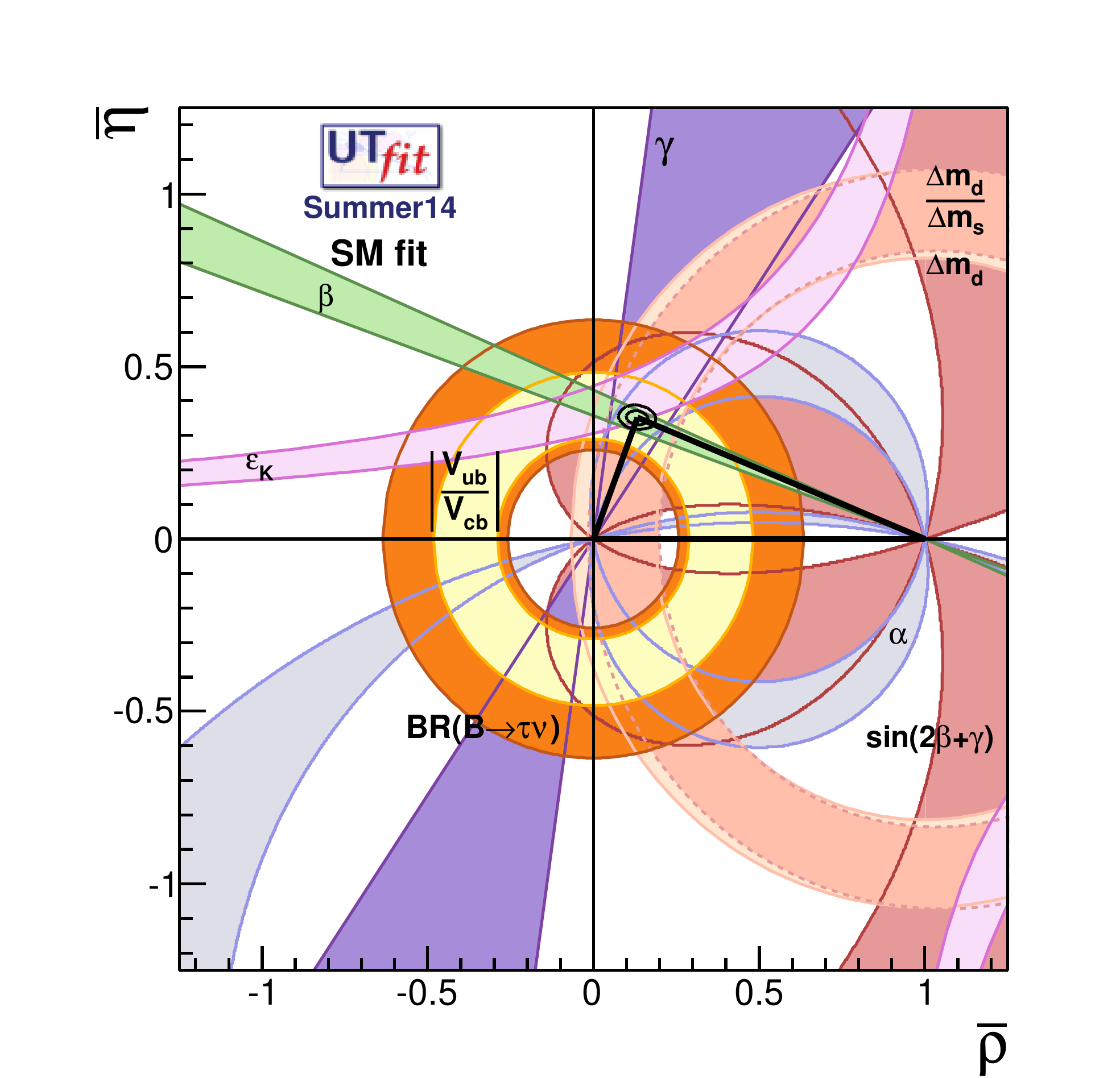}
   \end{center}
  \caption{Latest constraints on the CKM matrix element expressed in the $(\overline{\rho}, \overline{\eta})$ plane.  From Ref.~\protect{\cite{CKMfitter}}. Color online.}
  \label{fig:CKM}
\end{figure}

The constraint on the CKM matrix is usually written on the complex $(\overline{\rho}, \overline{\eta})$ plane defined by 
\begin{eqnarray}
\overline{\rho}+i \overline{\eta}=\left(1-\frac{\lambda^2}{2}\right)\left(\rho+i\eta\right).
\end{eqnarray}
In Fig.~\ref{fig:CKM} we show the latest result compiled by the  UTfit  Group~\cite{UTfitter}.  The result by the CKMfitter group~\cite{CKMfitter} is similar.  The inside of various regions are allowed from each of the constraints.  There has to be a common overlap region if the Standard Model is to be consistent with experiment.  Fitting data to the Standard Model, UTfit finds  
\begin{eqnarray}
\overline{\rho}=0.132(23), \quad \overline{\eta}=0.351(13).
\end{eqnarray}

\begin{table}[b]
\caption{CP violating observables and relevant matrix elements.  Percentage values of errors are also listed. See text for explanations.}
\label{tab:CKM}  
\begin{tabular}{llllll}
\hline\noalign{\smallskip}
observable & PDG~\cite{PDG}/ &    & matrix & FLAG~\cite{FLAG2014}&   \\
		  &HFAG~\cite{HFAG} &	& element		&			&	\\
\noalign{\smallskip}\hline\noalign{\smallskip}
$\epsilon$  & $2.228(11)\times 10^{-3}$&0.49\%&$\hat{B}_K$&0.7661(99)&1.3\%\\
$\Delta m_s/\Delta m_d$&$34.8(2)$&0.60\%&$\frac{f_{B_s}\sqrt{B_{B_s}}}{f_{B_d}\sqrt{B_{B_d}}}$&$1.268(63)$~\cite{FNAL2012}&5.0\%\\
${\cal B}(B\to\tau\nu)$&$1.14(27)\times 10^{-4}$&24\%&$f_B ({\rm MeV})$&$190.5(4.2)$&2.2\%\\
${\cal B}(B\to\pi\ell\nu)^{1)}$&$0.38(2)\times 10^{-4}$&5\% & $\Delta\zeta^{B\to\pi} $& 2.16(50)~\cite{HPQCD2006,FNAL2008} & 23\%\\
$B\to D^*\ell\nu$ & $35.85(45)\times 10^{-3}$& 1.3\% & $F^{B\to D^*}(1)$& 0.906(13)~\cite{FNAL2014} & 1.4\%\\
\noalign{\smallskip}\hline
\multicolumn{4}{l}{${}^{1)}$ integral over $16 {\rm GeV}^2\le q^2\le q_{\rm max}^2$}\\
\end{tabular}
\end{table}

For each region in Fig.~\ref{fig:CKM} the width of the allowed band represents experimental uncertainties as well as those of lattice QCD determinations.  In the two left columns of Table~\ref{tab:CKM}, we list the experimental inputs, and their percentage errors,  used for constraining the allowed regions.  The quantities and percentage errors listed in the two right columns are the QCD matrix elements which are needed to convert the experimental inputs to constraints in the $(\overline{\rho},\overline{\eta})$ plane.  

The experimental values are taken from Reviews of Particle Physics~\cite{PDG} and a compilation of heavy flavor data by the Heavy Flavor Averaging Group~\cite{HFAG}.  
The lattice QCD values for hadronic matrix elements are taken from a recent compilation by the Flavor Lattice Averaging Group (FLAG)~\cite{FLAG2014}.  We select the values quoted for $N_f=2+1$ calculations that passes FLAG criteria for the control of systematics errors including the continuum extrapolation.  In some cases, only a few  calculations are available, in which cases we quote the original references.  

We observe that B meson related quantities still has a significant margin for improvement, both on the experimental and lattice QCD sides.  The SuperKEKB experiment, which will start in a few years, is expected to reduce the experimental error by another order of magnitude.  Improvements in  lattice QCD values should keep pace. 
We should also remark that the large error in the $B$ decay measurement affects the determination of $V_{cb}$, which in turn broadens the band for $\epsilon$.  The determination of $B_K$ itself is already quite precise. 

\subsubsection{CP violation in the two pion decays of K meson}

Historically, CP violation was discovered through an observation of K mesons decaying into two pions in 1964.  The strength of direct CP violation relative to that in the state mixing is measured by the ratio
\begin{eqnarray}
\frac{\epsilon'}{\epsilon} = \frac{\omega}{\sqrt{2}\vert\epsilon\vert} \left[\frac{{\rm Im}A_2}{{\rm Re}A_2}-\frac{{\rm Im}A_0}{{\rm Re}A_0}\right],
\end{eqnarray}
where $A_I$ denotes the decay amplitude with isospin $I$ in the final state, and 
\begin{eqnarray}
\omega^{-1}=\frac{{\rm Re}A_0}{{\rm Re}A_2}.
\end{eqnarray}
Two heroic experiments, NA48 at CERN~\cite{NA48} and KTEV at FNAL~\cite{KTeV}, spanning two decades from the 80's to early 2000's,  measured $\epsilon'/\epsilon$ with the result,
\begin{eqnarray}
\frac{\epsilon'}{\epsilon} =\left\{\begin{array}{ll}
						18.5(7.3)\times 10^{-4},&\quad {\rm NA48}\\
						&\\
						20.7(2.8)\times 10^{-4},&\quad {\rm KTeV}\\
						\end{array}
						\right.
\end{eqnarray}
It has also been a long standing puzzle that the amplitude for the $I=0$ final state is sizably larger than that for $I=2$, namely $\omega^{-1}\approx 22$.  This is called the $\Delta I=1/2$ rule.  Whether the Standard Model can successfully explain these features of $K\to 2\pi$ decay has been a major problem in particle physics.  Since the issue boils down to calculating the strong interaction corrections to the effective weak interaction Hamiltonian,  this has been an important challenge to lattice QCD since the 80's. 

There are three obstacles to a successful calculation of the $K\to\pi\pi$ amplitudes in lattice QCD.   The first obstacle is chiral symmetry. Chirality plays an essential role in weak interactions.  The effective 4-quark weak interaction Hamiltonian obtained at a lattice cutoff scale of a few GeV starting from a much higher weak interaction scale has a definite chiral structure~\cite{Buras}.  Thus one has to employ a lattice fermion formulation which has chiral symmetry.  If, on the other hand, one uses non-chiral formulations such as Wilson's, one has to carefully control chiral symmetry violation effects under renormalization.  The former option was not available until the late 90's when domain-wall and overlap formulations were proposed.   The latter problem was successfully resolved for the Wilson fermion action only recently, with an enticing conclusion that the renormalization structure is the same as in the continuum for $K\to\pi\pi$ decay~\cite{Doninietal1999,Ishizuka2013}. 

The second obstacle stems from the fact that,  in the Euclidean Green's function for the $K\to\pi\pi$ transition by the weak Hamiltonian, the two-pion state with zero relative momentum, being the state with lowest energy in this channel, dominates for large times.  This contradicts the physical kinematics of the decay; the two pions decaying from a K meson at rest should have an equal and opposite momentum $p=\sqrt{m_K^2/4-m_\pi^2}$.  Thus a naive calculation does not yield physical results.  A resolution was found in 2001~\cite{LelloucheLuscher}.  One makes a calculation for a finite lattice volume chosen such that the energy of the two pion state matches the energy of the K meson.  The physical amplitude for infinite volume can then be obtained by the following formula,
\begin{eqnarray}
\vert A_{phys}(K\to\pi\pi)\vert^2=8\pi\frac{E_{\pi\pi}}{p}\left\{p\frac{d\delta_{\pi\pi}(p)}{dp}+q\frac{d\phi(q)}{dq}\right\}\vert\left<K\vert H_W\vert\pi\pi\right>_{lat}\vert^2.
\end{eqnarray}
Here $p$ is the pion momentum, $\delta_{\pi\pi}(p)$ the elastic $\pi\pi$ phase shift  at momentum $p$, $q=pL/(2\pi)$, and $\phi(q)$ is defined by 
\begin{eqnarray}
\tan\phi(q)=-\frac{q\pi^{3/2}}{Z_{00}(1;q^2)}, \quad Z_{00}(1;q^2)=\frac{1}{\sqrt{4\pi}}\sum_{\vec{n}\in Z^3}\frac{1}{\vec{n}^2-q^2}.
\end{eqnarray}

The third issue is specific to the $I=0$ channel and computational in character.    In this channel, there are diagrams with disconnected quark loops, {\it e.g.,}  quarks from the pions annihilate themselves.  
In addition, the so-called Penguin diagrams in which a pair of quarks from the weak Hamiltonian forms a loop are also present.  These diagrams suffer from large statistical fluctuations,  rendering the statistical average difficult.  This problem is gradually being overcome with efficient algorithms for computing disconnected and Penguin contributions, and with increase of computing power which allows a large number of Monte Carlo ensembles. 

Recently there has been significant progress assembling these developments together.  The RBC Collaboration has developed applications of the domain-wall formulation of QCD having chiral symmetry.  It has succeeded in calculating the $I=2$ amplitude for the physical pion mass~\cite{RBRC2012}.  Their result, obtained at a lattice spacing of $a\approx 0.14$~fm,  is given by 
\begin{eqnarray}
{\rm Re}A_2&=&+1.381(46)_{\rm stat}(258)_{\rm syst}\times 10^{-8} {\rm GeV},\\
{\rm Im}A_2&=&-6.54(46)_{\rm stat}(120)_{\rm syst}\times 10^{-13} {\rm GeV}.
\end{eqnarray}
The real part is in good agreement with experiment: ${\rm Re}A_2^{\rm exp}=1.479(4)\times 10^{-8}$~GeV. 

The RBC Collaboration has been attacking the much more difficult $I=0$ channel, using a special $G$-periodic boundary condition to force the pions to carry momentum so that an energy matching is realized between the K meson and 2 pions.  Preliminary results have been presented at Lattice 2014 Conference this year.  

Another group at Tsukuba has been pursuing the problem using the Wilson fermion formulation~\cite{Ishizuka2013}.  The renormalization structure for the operator relevant for $K\to\pi\pi$ decay turned out to be the same as in the continuum except for the mixing with dimension 3 operator $\overline{s}\gamma_5d$, which is subtracted non-perturbatively.  Their calculation so far does not achieve physical kinematics.  The $\pi$ meson mass is artificially taken large so that $2m_\pi=m_K$ is satisfied.  Nonetheless they reported an encouraging first result for $I=0$ channel as well as for $I=2$ at Lattice 2014 Conference.  

It is hoped that progress in the near future brings definitive results on the $I=0$ amplitude $A_0$.  This will put an independent constraint on the CP violating phase $\bar{\eta}$ in Fig.~\ref{fig:CKM} as 
the ratio $\epsilon'/\epsilon$ is proportional to $\bar{\eta}$. 

\subsection{Quark-gluon matter at high temperature and density}
\label{sec:3-4}

Confinement of quarks and spontaneous breakdown of chiral symmetry are both dynamical consequences of QCD.  A very interesting question then is how these properties may or may not change if parameters (or dials) external to QCD are varied.  One of important dials is temperature, which increases toward the Big Bang in the Early Universe. 
Another dial is baryon number density, which has a large value in extreme conditions such as in the core of neutron stars.

In both cases one is interested in an aggregate of hadrons,  rather than individual hadrons, and wants to understand how its properties change at extremely high temperature or density.  As we discuss below, a general conclusion from lattice studies is that a gas of hadrons turns into a different state in which quark and gluon degrees of freedom becomes manifest.  The novel state of matter is often called quark-gluon plasma (QGP).  

\subsubsection{Phase diagram of QCD at finite temperature:analytical considerations}
\label{sec:3-4-1}

The Euclidean formulation adopted for lattice QCD is well suited for studies of its properties at finite temperatures.  If one considers a lattice with $N_t$ sites in the temporal direction,  and imposes periodic and antiperiodic boundary condition for gluon and quark fields, respectively, the lattice path integral (\ref{eq:partition}) is equal to the canonical  partition function 
\begin{eqnarray}
Z_{\rm QCD}={\rm Tr}\left[\exp\left(-H_{\rm QCD}/T\right)\right],
\end{eqnarray}
at a physical temperature  
\begin{eqnarray}
T=\frac{1}{N_ta}.
\end{eqnarray}
This connection shows that methods developed for zero temperature studies, including Monte Carlo calculations,  are readily applicable to the finite temperature case. 

\begin{figure}
\begin{center}
  \includegraphics[bb= 0 0 612 792,width=0.45\textwidth]{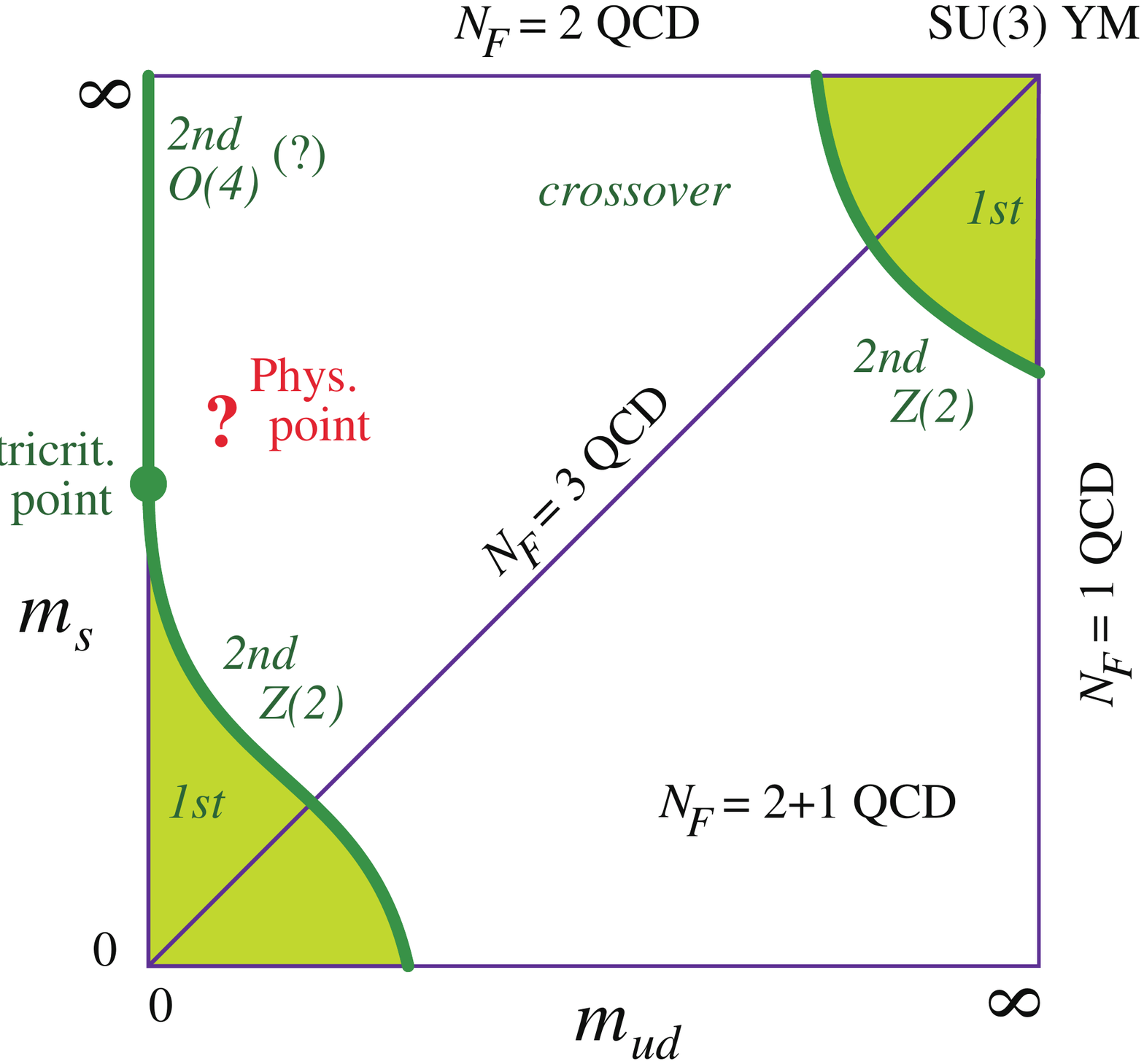}
  \hspace{7mm}
  \includegraphics[bb= 0 0 612 792,width=0.45\textwidth]{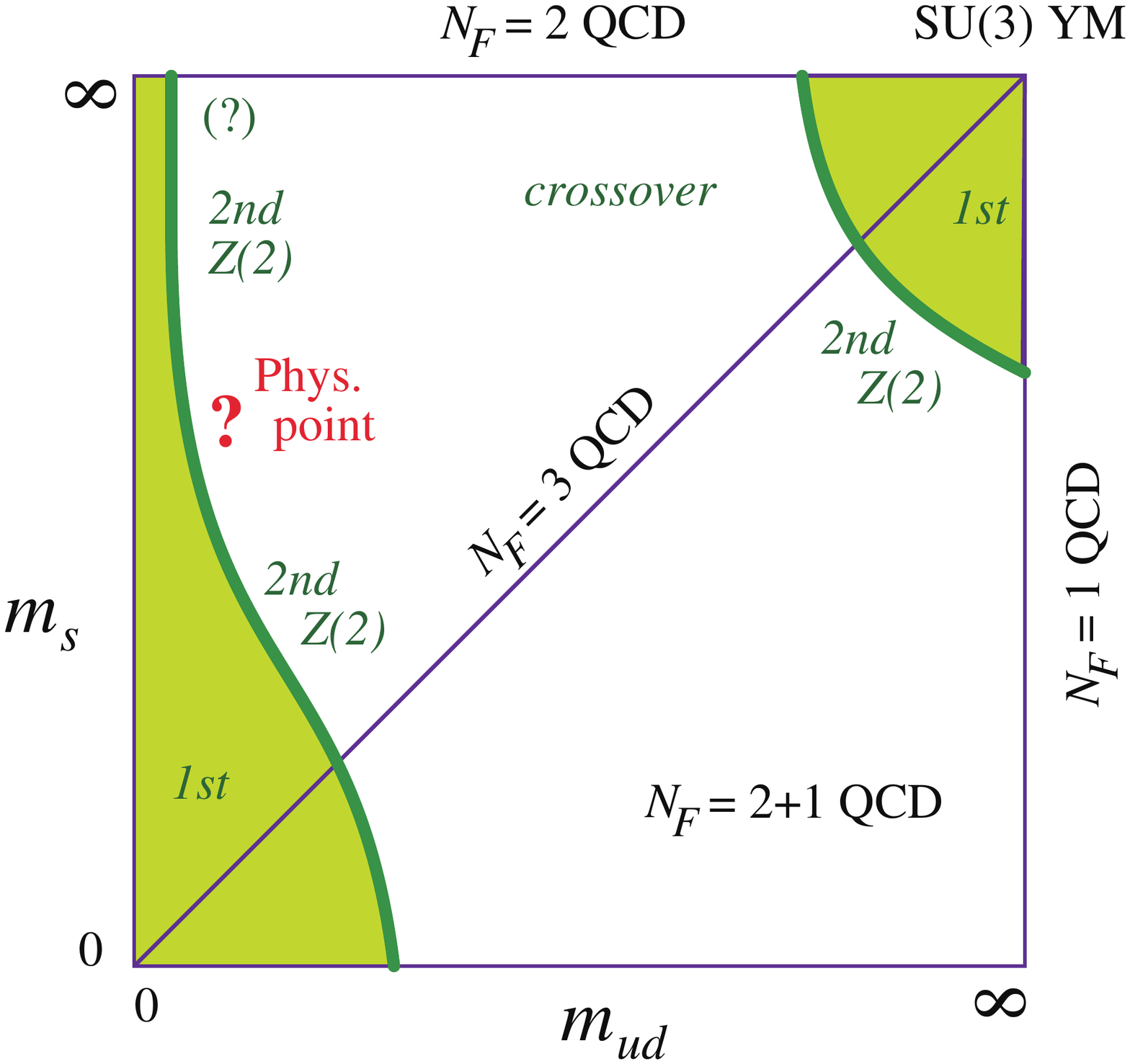}
  \vspace*{-17mm}
\end{center}
\caption{Phase digram in 2+1 flavor QCD as a function of the degenerate $u$ and $d$ quark mass $m_{ud}$ and the $s$ quark mass $m_s$. Left panel shows the case for a second order chiral transition for two-flavor QCD.   Right panel shows the case for a first order  two-flavor chiral transition. From Ref.~\cite{Kanaya2010}.  Color online.}
\label{fig:columbia}
\end{figure}

In order to discuss what happens as the temperature is raised from zero, it is helpful to consider a two-dimensional plane of the average up-down quark mass $m_{ud}$ and the strange quark mass $m_s$, often dubbed Columbia Plot, as depicted in Fig.~\ref{fig:columbia}~\cite{Columbia1990}.  One can examine various limiting cases as follows.  

\paragraph{(i) Pure gluon theory:} 
The top right corner in Fig.~\ref{fig:columbia} for $m_{ud}=m_s=\infty$ corresponds to the pure gluon theory without quark degrees of freedom.  This case is important, nonetheless, since confinement is a dynamical consequence of the gluon fields.  

The pure gluon theory possesses a center  $Z(3)$ symmetry defined by the transformation
\begin{eqnarray}
\label{eq:centerZ(3)}
U_{n4}\to \zeta U_{n4}, \quad \zeta\in Z(3),
\end{eqnarray}
for the sites $n$ on some fixed time slice $t$.   The corresponding order parameter is the Wilson loop winding around the space-time in the time direction at a fixed spatial site $\vec{n}$ defined by 
\begin{eqnarray}
\label{eq:polyakov}
\Omega(\vec{n}) =  {\rm Tr}\left(\prod_{n_t=1}^{N_t}U_{(\vec{n},n_t)4}\right).
\end{eqnarray}
This operator, often called Polyakov loop,  transforms under the center Z(3) symmetry as  
\begin{eqnarray}
\Omega(\vec{n}) \to \zeta \Omega(\vec{n}).  
\end{eqnarray}
Polyakov\cite{Polyakov} and Susskind~\cite{Susskind} argued that the center Z(3) symmetry is intact at low temperatures with 
$\left<\Omega(\vec{n})\right>=0$, but becomes spontaneously broken beyond a certain temperature $T=T_c$ with a non-zero expectation value $\left<\Omega(\vec{n})\right>\ne 0$.   

The connection with deconfinement is most easily understood in the following way.  The 2-point correlation function of Polyakov loops defined by 
$\left<\Omega(\vec{n}) \Omega^\dagger(\vec{m})\right>$ is connected with the free energy 
$F(\vert\vec{n}-\vec{m}\vert)$ of a static quark at site $\vec{n}$ and an antiquark at site $\vec{m}$  by 
\begin{eqnarray}
\exp\left(-F(r)/T\right) = \left<\Omega(\vec{n}) \Omega^\dagger(\vec{m})\right>, \quad r=\vert\vec{n}-\vec{m}\vert.
\end{eqnarray}
Assuming the presence of a mass gap $\mu$, the 2-point function on the right hand side is expected to behave for large $r$ as 
\begin{eqnarray}
\left<\Omega(\vec{n}) \Omega^\dagger(\vec{m})\right>
\to  
\left<\Omega(\vec{n})\right>\cdot \left< \Omega^\dagger(\vec{m})\right> + O\left( \exp\left(-\mu r\right) \right),\quad r\to\infty.
\end{eqnarray}
Depending on the expectation value $\left<\Omega\right>$, one finds for $r\to\infty$ that  
\begin{eqnarray}
 F(r) &\to& \left\{\begin{array}{ll}
						\sigma r,    &\quad \left<\Omega\right>=0, \quad T<T_c \\
						&\\
						c + O(\exp\left(-\mu_D r\right)), &\quad \left<\Omega\right>\ne 0, \quad T>T_c\\
						\end{array}
						\right.
\end{eqnarray}
where $\sigma=T\mu$ for $T<T_c$,  and $c$ is a constant and $\mu_D=\mu$ is a color electric Debye screening mass for $T>T_c$. 

Analytical considerations indicate that this deconfinement phase transition should be of first order~\cite{YaffeSvetitsky}.  If this is the case, the first-order transition will extend into the region of large but finite quark masses~\cite{BanksUkawa}, terminating at a line of second-order transition as depicted in Fig.~\ref{fig:columbia}. 

\paragraph{(ii) $N_f=2$ and $N_f=3$ massless QCD:}
The top left corner of Fig.~\ref{fig:columbia} corresponds to $N_f=2$ massless QCD with $m_{ud}=0$ but $m_s=\infty$, and the bottom left corner to $N_f=3$ massless QCD with $m_{ud}=0$ and $m_s=0$. 

For $N_f$ flavors of massless quarks, QCD is invariant under a global $SU(N_f)\otimes SU(N_f)$ chiral symmetry, which is spontaneously broken down to vector $SU(N_f)$ symmetry with a non-zero vacuum expectation value of the order parameter $\Sigma=\left<\overline{q}_nq_n\right>\ne 0$. 
The $N_f^2-1$-plet of pseudo scalar mesons are the corresponding  Nambu-Goldstone bosons.  

When one raises the temperature, thermal fluctuations tend to destabilize the chiral order parameter $\Sigma$.  Thus one expects a phase transition restoring chiral symmetry at some temperature.  A more detailed examination is possible using an effective non-linear sigma model for the order parameter field $\Phi_n^{ij}=\overline{q}_n^iq_n^j$ with $i,j=1, \cdots, N_f$ and renormalization group methods.  The result~\cite{PisarskiWilczek} indicates that (i) for $N_f=3$ or larger, chiral symmetry is restored through a first order phase transition, while (ii) for $N_f=2$ the phase transition is of first or of second order depending on whether flavor singlet $U(1)$ axial symmetry is effectively restored or not at $T>T_c$.   

\paragraph{(iii) Connecting $N_f=2$ and $N_f=3$ massless QCD:} 
One can interpolate between the $N_f=2$ and $3$ cases by changing the strange quark mass $m_s$ from $\infty$ to $0$.  If the $N_f=2$ transition is of second order, the first order transition for $N_f=3$ at $m_s=0$ has to change to a second order transition at a tricritical point at some $m_s=m_s^c$. 

\paragraph{(iv) $N_f=3$ symmetric QCD:}
In general, a first order phase transition is stable under symmetry breaking perturbations up to a critical value where it terminates with a second order phase transition.  
For the second order case, the phase transition immediately disappears if symmetry breaking perturbations are turned on.  Thus one expects a sheet of first order transition extending from the bottom left corner corresponding to $N_f=3$ massless QCD to the interior of the phase diagram.  The sheet should terminate at a line of second order phase transitions, which is expected to belong to the Ising universality class~\cite{GavinGokschPisarski}.  

If the $N_f=2$ transition at $m_{ud}=0$ is of second order, this line of second order transitions will hit the vertical axis at $m_s=m_s^c$ with a power law $m_{ud}\propto (m_s^c-m_s)^{5/2}$~\cite{Rjagopal1995}, as depicted in the left panel of Fig.~\ref{fig:columbia}.
 If, on the other hand, the $N_f=2$ transition is of first order, the line will go up to the top horizontal line (the right panel of Fig.~\ref{fig:columbia}).  

\subsubsection{Monte Carlo study of the finite-temperature phase diagram}
\label{sec:3-4-2}

Lattice QCD simulations are carried out for a fixed temporal lattice size $N_t$.  One then regulates the temperature indirectly by varying the bare gauge coupling constant $g_0^2$; since the continuum limit $a=0$ is located at $g_0^2=0$, weaker couplings correspond to smaller lattice spacing, and hence to higher temperatures, and {\it vice versa} for stronger couplings and to lower temperatures.  

The basic tool for studying the phase diagram is finite size scaling theory developed in the late 60's and 70's~\cite{FSS}.  This theory helps to analyze how singularities marking phase transitions develop from numerical data obtained at finite volumes. 

\paragraph{(i) Pure gluon theory:} The first instance where finite size scaling method was crucially effective was the deconfinment transition of the pure gluon theory.  Early simulations quickly found that the Polyakov loop expectation value exhibited a rapid increase from $\left<\Omega\right>\approx 0$ to non-zero values over a narrow range of temperature as expected.  A subtle issue is if a rapid increase actually signifies a phase transition, and if so, whether it is a first order phase transition with a discontinuous jump of $\left<\Omega\right>$ at $T_c$ or a second order phase transition with a continuous $\left<\Omega\right>$ but having a singular  derivative.  

The susceptibility of Polyakov loop is defined by 
\begin{eqnarray}
\chi_\Omega=\frac{1}{L^d}\sum_{\vec{n},\vec{m}}\left<\Omega(\vec{n})\Omega^\dagger(\vec{m})\right>
\end{eqnarray}
with $L$ the linear extent of the system, and $d$ the space dimension.  For a finite volume, the susceptibility exhibits a peak.  The infinite behavior of the peak height $\chi_\Omega^{\rm max}$ distinguishes the order of the phase transition.  According to finite size scaling theory, if one parametrizes the volume dependence by a power law, 
\begin{eqnarray}
\chi_\Omega^{\rm max}(L)\propto L^\alpha, \quad L\to\infty,
\end{eqnarray}
the value $\alpha=d$, {\it i.e.,} the dimensionality of space, signals a first order transition, while values $\alpha<d$ characterizes a second order transition in which case $\alpha$ is related to the critical exponents.  Similar criterion holds for the volume dependence of the width of the peak.  

These criteria were utilized to establish the first order nature of the deconfinement transition in the pure gluon theory~\cite{FukugitaOkawaUkawa}, in agreement with the analytical considerations. 

\paragraph{(ii) Chiral transition with quarks:} The chiral phase transition in the presence of quarks has also be studied extensively.  Ideally one would like to use a fermion formulation preserving chiral symmetry such as the domain-wall or overlap.  They became available only in the early 2000's, and are computationally very costly, however.  For these reasons, most of the calculations have utilized, and still use,  the staggered fermion formulation and, to a lesser extent, the Wilson formulation.  

Broadly speaking, results accumulated to date are as follows.  For $N_f=3$ degenerate flavors, the phase transition is consistent with being of first order for small quark masses.  Increasing the quark mass, the transition weakens and terminates at a second order transition whose exponents are consistent with the Ising universality class  (see Refs.~\cite{Kayaetal1999,Karschetal2001} for early representative work and references).   These calculations are made at most for $N_t=6$ lattices, and the continuum limit is yet to be taken~\cite{Dingetal2011}.  

For $N_f=2$, the situation is more complicated.  Early simulations were consistent with a second order phase transition both with the staggered~\cite{JLQCDNf2Thermo}  and with the Wilson~\cite{CP-PACSThermoNf2001} fermion formulations.  For the staggered case, however, the critical exponents did not come out consistent with either $O(4)$ nor $O(2)$ values.  There have also been simulations suggesting  consistency with a first order transition~\cite{Bonatietal2009}.  

As already pointed out in Ref.~\cite{PisarskiWilczek}, the order of the $N_f=2$ chiral transition is connected with the $U_A(1)$ anomaly.  Recently, a theoretical argument has been put forward~\cite{AokiFukayaTaniguchi}  that the anomaly effects disappear for certain sets of correlation functions if chiral symmetry is restored.  At the moment, the order of the $N_f=2$ transition is not settled completely.

\subsubsection{Thermodynamics with physical quark masses}
\label{sec:3-4-3}
 
Physically, the crucial issue is where the point with physical quark masses lies on the phase diagram in Fig.~\ref{fig:columbia}.  The staggered results are unanimous that it lies beyond the line of critical end points.  Hence there is only a continuous crossover and no phase transition with a singular behavior.  The basis for this conclusion includes an extensive study at physical quark masses with infinite volume and continuum extrapolations using susceptibilities of various physical observable~\cite{WB2006}.  

Since the transition is a continuous crossover, the transition temperature is not uniquely determined but depends on the quantity used.  For example~\cite{WBTc2010}, one finds $T_c=147(2)(3)$~MeV if one uses the susceptibility peak of chiral order parameter, and $T_c=157(4)(5)$~MeV from the inflection point of the energy density.  An independent calculation~\cite{HotQCD2012} reported $T_c=154(9)$~MeV from $O(4)$ scaling analysis of chiral susceptibility.  The results are consistent, and altogether indicate $T\approx 150-160$~MeV as the temperature range across which the physical chiral transition takes place. 

\begin{figure}[t]
\begin{center}
  \includegraphics[bb=0 0 411 428,width=0.45\textwidth]{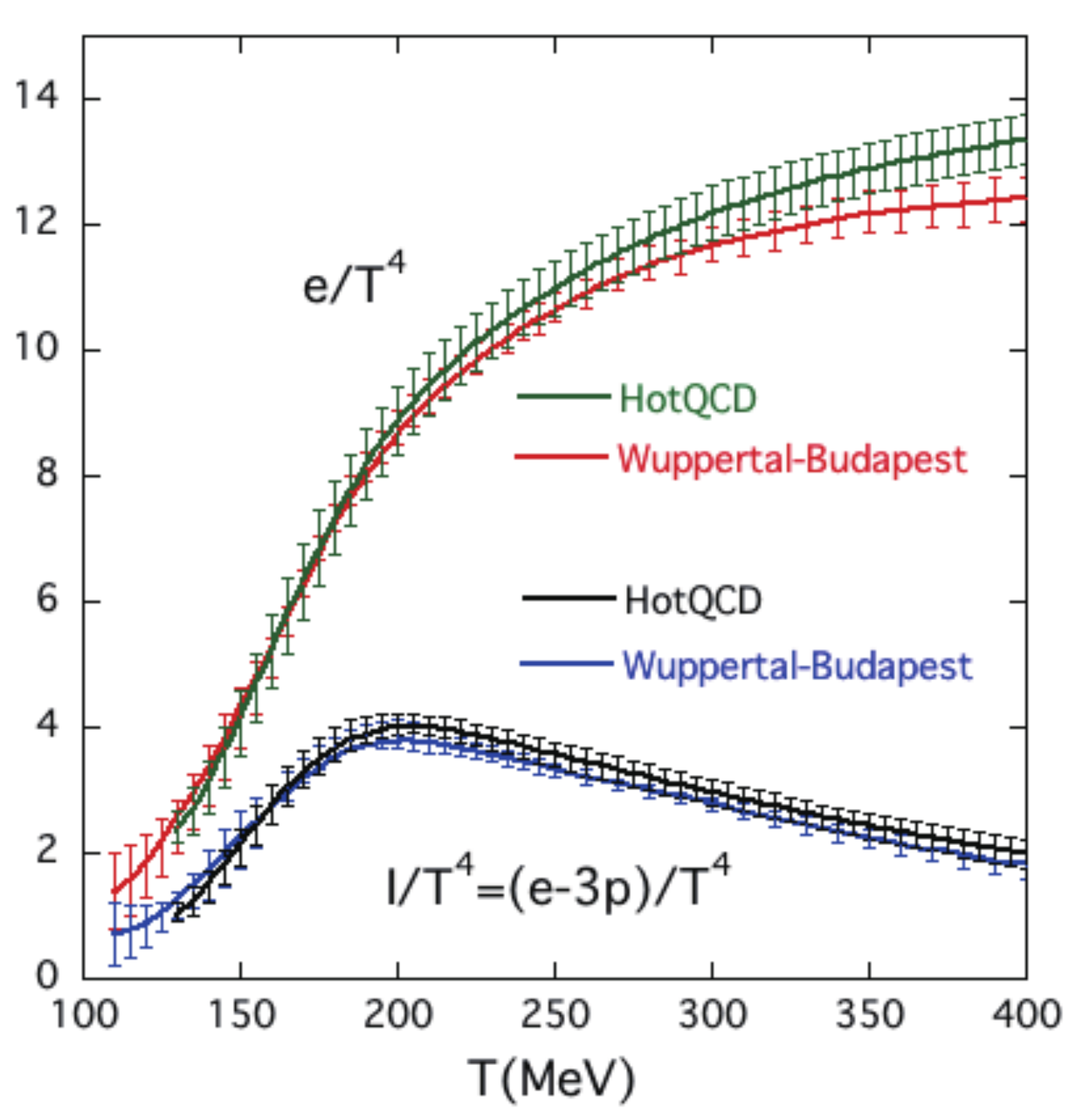}
  \hspace{7mm}
  \includegraphics[bb=0 0 424 433,width=0.46\textwidth]{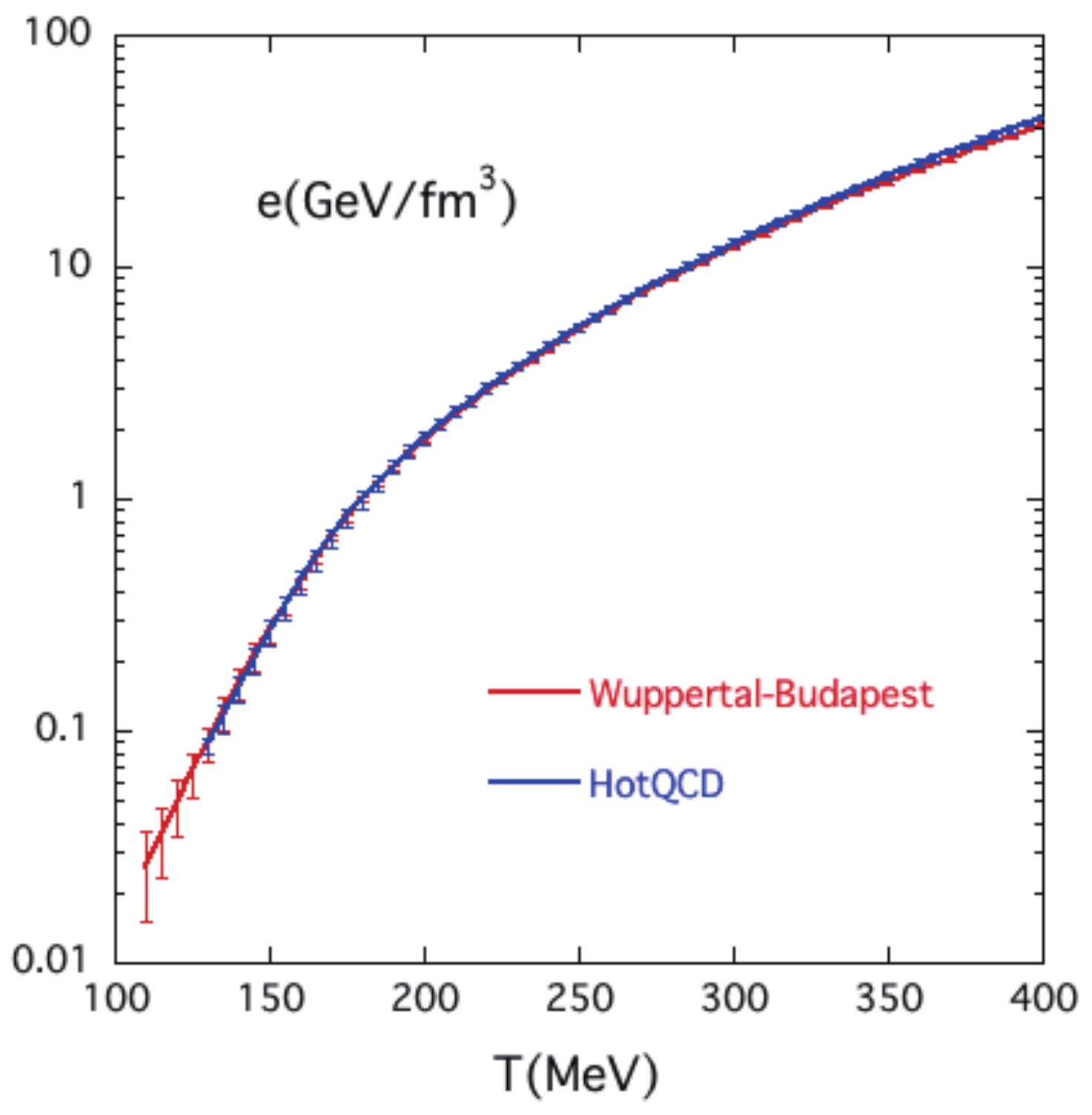}
\end{center}
\caption{Thermodynamic quantities in the continuum limit in 2+1 flavor QCD as a function of temperature.  Left panel shows the energy density $e/T^4$ and interaction measure $I/T^4=(e-3p)/T^4$ in units of $T^4$, and right panel shows energy density $e$ in units of GeV/fm$^3$.  From Refs.~\protect{\cite{WBEoS2014,HotQCD2014}}. Color online.}
\label{fig:thermo}
\end{figure}

In the left panel of Fig.~\ref{fig:thermo} we show the energy density $e/T^4$ and interaction measure $I/T^4=(e-3p)/T^4$, with $p$ the pressure, in units of $T^4$ calculated in the staggered quark formalism by two groups~\cite{WBEoS2014,HotQCD2014}.  They are obtained at the physical quark masses,  and infinite volume and continuum extrapolations are made.  We observe very good agreement up to about 200~MeV beyond which a one standard deviation difference appears.   Physically important is the feature that the Stefan-Boltzmann value for free gluons and quarks, $e_{\rm SB}/T^4=\pi^2(8+21 N_f/4)/15=15.62\cdots$ for $N_f=3$, is reached only slowly, with significant deviations remaining at $T/T_c\sim 2 - 3$.  This indicates that the quark-gluon matter  is strongly interacting at these temperature ranges.   The right panel shows the energy density in units of GeV/fm${}^3$. 

Experimental effort toward detection of quark-gluon plasma through heavy ion collisions in accelerators has been going on for a long time.  It started  with the Bevalac at Berkeley in the 70's, the AGS at BNL and the SPS at CERN in the 80's, with RHIC at BNL since 2000 and with the LHC at CERN most recently. 

Let us see how lattice QCD predictions compare with heavy ion experiments at RHIC and the LHC.  Phenomenological estimates on the energy density reached at the initial stage of the collision range from $e\approx 5.6$~GeV/fm$^3$ for Au--Au collisions at RHIC with $\sqrt{s_{NN}}=200$~GeV~\cite{PHENIX2005},  to $e\approx 15$~GeV/fm$^3$ for Pb--Pb collisions at the LHC with $\sqrt{s_{NN}}=2.76$~TeV~\cite{CMS2012}.   Looking at the right panel of Fig.~\ref{fig:thermo} we read out a temperature of order $T\approx 200$~MeV and $300$~MeV, respectively, for these energy densities which are high enough for the collision product to be in the high temperature phase.  For comparison, an experimental estimate of the temperature using the low $p_T$ excess of direct photons that are supposed to come from the initial thermalized state gave $T=221(27)$~MeV at RHIC~\cite{PHENIX2010} and $T=304(51)$~MeV at the LHC~\cite{ALICE2012}.  

The initial fireball rapidly cools as it expands, and hadrons are formed once the temperature falls below the transition temperature $T_c$.  It has been known that the yield of various hadrons from the collision could be well fitted with statistical thermal distribution parametrized by a chemical freeze out temperature $T_f$.  This temperature increases with the collision energy and saturates at $T_f\approx 160$~MeV at RHIC energies~\cite{Andronic2009}, though slightly decreasing at LHC energies.  

Furthermore, the azimuthal anisotropy in the transverse hadron yields, quantified in terms of the elliptic flow $v_2$ and  higher moments where $v_n=\langle \cos n\phi\rangle$, is well described by hydrodynamical models with no or small viscosity and using the equation of state from lattice QCD.

Thus, while small viscosity was a surprise, the experimental data are broadly consistent with the production of strongly interacting high temperature quark-gluon matter with the properties as predicted by lattice QCD.  It is interesting to note that small shear viscosity close to the quantum limit $\eta/s=1/4\pi$~\cite{Kovtunetal2005}  is observed by lattice QCD calculations of transport coefficients in the pure gluon case~\cite{NakamuraSakai2006,Meyer2007}.

From a heavy-ion collision point of view, the transition temperature and equation of state are only indirectly reflected in the characteristics of the final hadronic state.  The moments of conserved charges such as electric charge $Q$, strangeness $S$, and baryon number $B$ provide interesting quantities which are calculable in lattice QCD and are directly observable in experiments.  Thus they have been attracting much interest recently~\cite{Stephanovetal1999,EjiriKarschRedlich2006,Karsch2012,Bazabovetal2012,Borsanyietal2014}.

\subsubsection{Dynamics at finite baryon number density}
\label{sec:3-4-5}

\begin{figure}
\begin{center}
  \includegraphics[bb=0 0 570 381,width=0.6\textwidth]{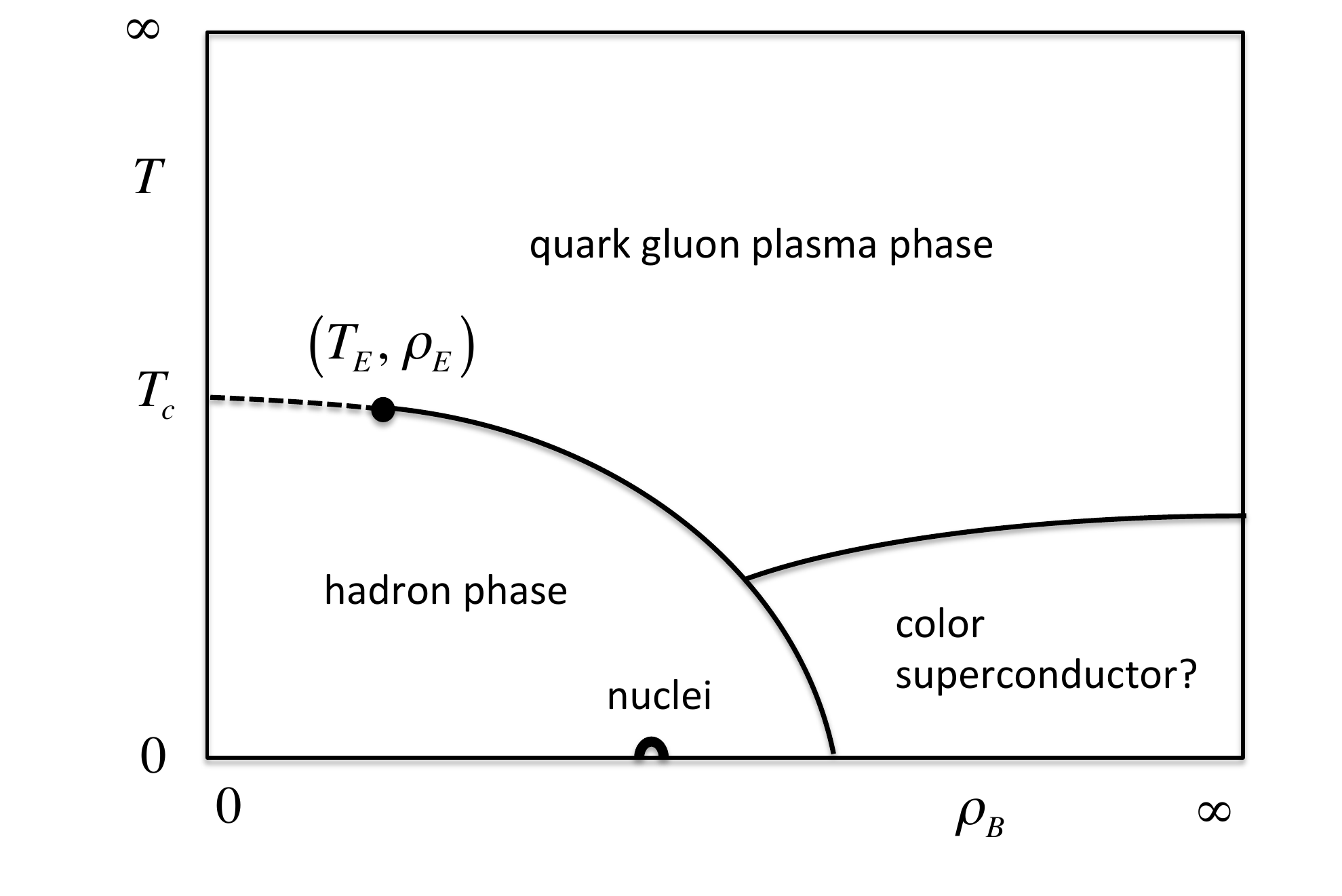}
\end{center}
\caption{Schematic phase diagram of QCD on the temperature baryon density plane.. }
\label{fig:Tmuplane}
\end{figure}

Theoretical expectations on the phases of QCD at finite temperature $T$ and finite baryon number density $\rho_B$ is depicted in Fig.~\ref{fig:Tmuplane}.  The chiral crossover transition at $T_c\approx 150-160$~MeV continues into the finite $\rho_B$ region, and hits a second order critical point at some $(T_E, \rho_E)$ beyond which the transition turns into a first order phase transition.  For sufficiently large baryon number density, one expects novel phases such as  a color superconductor. 

Finite baryon number density can be introduced by adding a quark chemical potential $\mu$ to the  Hamiltonian.  On a lattice, this is achieved by multiplying the positive temporal hopping term of the quark action by $\exp(\mu a)$, and the negative hopping term by $\exp( -\mu a)$.  

Solid quantitative results are still not available on the phase diagram. The main reason is that for non-zero quark chemical potential, the quark Dirac determinant $\det D(U, \mu)$ is complex, so that the Monte Carlo methods based on a probability interpretation of the weight $\det D\cdot \exp (-S_G(U))$ no longer work in general.  

One can see the difficulty by defining the phase of the determinant by 
\begin{eqnarray}
\theta (U, \mu)=-i\log\frac{\det D(U,\mu)}{\vert\det D(U,.\mu)\vert},
\end{eqnarray}
and rewriting expectation values in the following way:
\begin{eqnarray}
\left< O \right> &=&\frac{\left< O \exp\left( i\theta (U,\mu)\right)\right>_{\vert\vert}}{\left<  \exp\left( i\theta (U,\mu)\right)\right>_{\vert\vert}},
\end{eqnarray}
where $\left<\cdot\right>_{\vert\vert}$ means the average with respect to the phase quenched determinant $\vert\det D(U,\mu)\vert$.   The quenched average of the phase factor in the denominator is a ratio of the two partition functions:
\begin{eqnarray}
\left<  \exp\left( i\theta (U,\mu)\right)\right>_{\vert\vert}&=&\frac{Z(\mu)}{Z_{\vert\vert}(\mu)},\\
Z_{\vert\vert}(\mu)&=&\int\prod_{n\mu}dU_{n\mu}\vert\det D(U,\mu)\vert\exp (S_{\rm gluon}(U)).
\end{eqnarray}
If the quenched average defines a statistical system with a free energy density $f_{\vert\vert}(\mu)$, one can write 
\begin{eqnarray}
\left<  \exp\left( i\theta (U,\mu)\right)\right>_{\vert\vert}=\exp\left(-\frac{V (f(\mu)-f_{\vert\vert}(\mu)}{T}\right),
\end{eqnarray}
with $f(\mu)$ the free energy density of the original system.  The right hand side vanishes exponentially fast for large spatial volumes $V\to\infty$.  This is one way of explaining the sign problem. 

A variety of methods have been devised and explored to overcome this problem.  Besides the phase quenched calculation described above, they include reweighting of the determinant~\cite{FodorKatz2001}, analytic continuation from imaginary chemical potential~\cite{RobergeWeiss1986,ForcrandPhilipsen2002}, Taylor expansion in powers of $\mu/T$~\cite{Alltonetal2003,Ejirietal2012}, canonical ensemble simulation~\cite{Lietal}, complex Langevin simulation~\cite{Aartsetal,Sexty2013}, and others.  These methods have yielded some success for not too large values of $\mu/T$.  The main problem, however,  has been the difficulty of controlling errors in the results.  The precise location of the critical point E and physics of  finite density QCD for larger baryon number density is still largely open at present.  

\section{Conclusions}
\label{sec:4}

Lattice QCD is a major contribution of Kenneth Wilson to physics.  In our view, it is  on a par in significance with his renormalization group theory of critical phenomena which won him Nobel Prize in Physics in 1982.   Conceptually, it clarified how  quantum fluctuations of gauge fields give rise to a confining force which is essentially distinct in its origin from the forces arising from exchange of a particle such as the electromagnetic force.  At the same time, coupled with supercomputers, it opened a way to calculate the physical predictions from its first principles, making possible detailed comparisons with experiment.  

In 2004, Kenneth Wilson delivered a talk entitled ``The Origins of Lattice Gauge Theory" at International Symposium on Lattice Field Theory held at Fermi National Accelerator Laboratory~\cite{Wilson2005}.  Looking back on the development of lattice QCD, he said that {\it ``The lattice gauge theory was a discovery waiting to happen, once asymptotic freedom was established."}, and went on to describe works of his contemporaries who could have preceded or shared the discovery.  Nonetheless  it is clear that, because of his previous studies leading to his encounter with statistical mechanics and phase transitions, he was in a unique position to be the first to grasp the deep significance of strongly coupled gauge dynamics in relation to confinement.    

In the same article, he commended on the vast progress in lattice QCD in the thirty years since 1974, the year of his seminal paper~\cite{Wilson1974}, due {\it ``in part to improved algorithms, in part to increased computing power, and in part to the increased scale of the research effort underway today"}, but urged on that {\it ``this does not mean that the present state of lattice gauge computations is fully satisfactory. The knowledge and further advances that will likely accumulate over the next thirty years should be just as profound"}.   

As if corroborating his view, in just a decade since then, the physical point computation was realized, and many beautiful physics results ensued, as summarized in Sec.~\ref{sec:3}.   

The past decade, then, was a turning point of lattice QCD in our  view; prior to this event, despite the premise that it provided a first-principle calculation, it remained a method of uncertain reliability, requiring chiral extrapolations which were difficult to control.    We are now able to calculate and understand many properties of hadron states including masses and matrix elements,  at the physical quark masses and controlling errors of calculations.  The precision of the calculation has now reached the level of a few \% or better in many quantities.  Consequently, important constraints are now available on the CKM matrix elements and CP violation in the Standard Model.  

This is not to say that progress has been uniform in all fronts.  In thermodynamics of QCD, while some basic quantities such as the transition temperature and equation of state have been calculated, the region of finite baryon number density is still largely unexplored.  A major methodological breakthrough will be needed to extend our understanding to the entire phase diagram of QCD. 

One area which did not even exist at the time of Wilson's talk in 2004 is nuclear physics based on lattice QCD.  In fact, while there were pioneering studies on H di-baryon in the late 80's~\cite{Mackenzie1985,Iwasaki1988} and nucleon scattering lengths in 1994~\cite{Fukugitaetal1994},  it is in 2007 that the two nucleon potential was first extracted from lattice QCD~\cite{Ishiietal2007}, and in 2009 that the binding energy of Helium was calculated directly from Helium correlation function ~\cite{Yamazakietal2009}.   This is a challenging area in terms of physics as well as in calculational techniques; one has to deal with a small energy scale of 0(10)~MeV, the number of Wick contractions for quark fields increases factorially fast with the mass number, and so does the statistical error of nuclei propagators for large time separations.  

In theoretical physics, theory and computing go hand in hand.  Calculations are indispensable in order to confirm the validity of theory, and even more so to explore the consequences of theory which help us better understand why our world works the way it does.   Lattice QCD is a prime example of such a relationship between theory and computing.  Kenneth Wilson clearly foresaw the importance and future potential of supercomputing in this connection, and many of his thoughts and vision in the early 80's~\cite{Lax1982} came to be realized since then.  Looking toward future, as he prophesied~\cite{Wilson1983}, our understanding of the strong interactions will become more profound as the computing power increases toward exascale and possibly beyond in the decades to come. 

\begin{acknowledgements}

My first encounter with lattice QCD was in the spring of 1974 in Tokyo, Japan, when  I was a graduate student under the supervision of Kazuhiko Nishijima at University of Tokyo.  Kenneth Wilson's preprint was introduced at a seminar organized by graduate students of the particle theory group.  I was struck by the novelty of the idea, which I vividly remember, but  I do not think I understood the full impact of that preprint at that time.  

I had a good fortune to spend two years from September 1975 to August 1977 as a postdoctoral fellow of Toichiro Kinoshita at Laboratory of Nuclear Studies of Cornell University.  Kenneth Wilson had his office one floor below.  John Kogut's office was on the same floor across the corridor, and there were a number of graduate students, all of them very active, including Gyan Bhanot, Paul Mackenzie, Michael Peskin, Serge Rudaz, Steve Shenker, and Junko Shigemitsu.   

At Cornell I worked with Tom Kinoshita on mass singularities and, in later terminology,  perturbative QCD.  I was of course aware of the work of Ken Wilson, operator product expansions, renormalization group, lattice QCD, Kondo problem, but somehow he was a somewhat distant figure for me throughout my stay at Cornell.  But I do remember him very well.  A lean and taciturn man, often with his shirt tail hanging out in the back, and he always wore a slight kindly smile on his face, which deepened from time to time when something apparently amused him.

My serious involvement with lattice QCD started after I left Cornell, first with analytic studies on Z(n) duality at Princeton, and then,  after I came back to Japan, with numerical simulations and even supercomputer development.   The 40 years of  lattice QCD since 1974 overlap with my scientific career.  It is an honor for me to write this article on Kenneth Wilson and lattice QCD. 
\\
\\
I would like to thank Sinya Aoki, Norman Christ, Carleton De Tar, Zoltan Fodor, Shoji Hashimoto, Yoichi Iwasaki, Kazuyuki Kanaya, Frithjof Karsch, Andreas Kronfeld,  Martin L\"uscher, and Paul Mackenzie for valuable comments on the manuscript. 

\end{acknowledgements}



%
%

\end{document}